\documentclass[12pt]{article}
\usepackage{amsmath,amsthm,amsfonts,amsbsy,amssymb}
\usepackage{url} 
\usepackage{graphicx,psfrag,epsf}
\usepackage{natbib}
\usepackage{url} 
\usepackage[english]{babel}
\usepackage[utf8]{inputenc}
\usepackage[T1]{fontenc}
\usepackage{algorithm} 
\usepackage{algpseudocode}
\usepackage{xcolor}
\usepackage{enumitem}
\usepackage{booktabs}
\usepackage{multirow}
\usepackage{colortbl}
\usepackage[colorlinks,citecolor=blue,urlcolor=blue,linkcolor=blue,linktocpage=true]{hyperref}

\newcommand{\N}{\mathbb{N}}

\newcommand{\R}{\mathbb{R}}

\newcommand{\E}[1]{\operatorname{\mathbb{E}}[#1]}

\newcommand{\monthword}[1]{\ifcase#1\or Jan\or Feb\or M\"ar\or Apr\or Mai\or Jun\or Jul\or Aug\or Sep\or Okt\or Nov\or Dez\fi}
\newcommand{\leadingzero}[1]{\ifnum #1<10 0\the#1\else\the#1\fi}             

\newcommand{\todayI}{\the\year"~\leadingzero{\month}"~\leadingzero{\day}}    	
\newcommand{\todayII}{\the\year\leadingzero{\month}\leadingzero{\day}}       	
\newcommand{\todayIII}{\leadingzero{\day}/\leadingzero{\month}/\the\year}    	
\newcommand{\todayIV}{\leadingzero{\day}.\leadingzero{\month}.\the\year}     	
\newcommand{\todayV}{\the\day.\the\month.\the\year}                          	
\newcommand{\todayVI}{\the\day.~\monthword{\the\month}. \the\year}           	
\newcommand{\todayVII}{\leadingzero{\day}.~\monthword{\the\month}. \the\year}	
\newcommand{\todayVIII}{\monthword{\the\month}. \the\year}										

\newcommand{\calD}{\mathcal{D}}

\newcommand{\calJ}{\mathcal{J}}

\newcommand{\calR}{\mathcal{R}}

\newcommand{\calT}{\mathcal{T}}

\newcommand{\mA}{\mathbf A}

\newcommand{\mB}{\mathbf B}
\newcommand{\vb}{\mathbf b}

\newcommand{\mH}{\mathbf H}

\newcommand{\mP}{\mathbf P}

\newcommand{\mW}{\mathbf W}

\newcommand{\mX}{\mathbf X}

\newcommand{\mY}{\mathbf Y}

\newcommand{\vbeta}{\mathbf \beta}

\makeatletter
\newcommand{\rd}{\@ifnextchar^{\DIfF}{\DIfF^{}}}
\def\DIfF^#1{%
   \mathop{\mathrm{\mathstrut d}}%
   \nolimits^{#1}\gobblespace}
\def\gobblespace{\futurelet\diffarg\opspace}
\def\opspace{%
   \let\DiffSpace\!%
   \ifx\diffarg(%
   \let\DiffSpace\relax
   \else
   \ifx\diffarg[%
   \let\DiffSpace\relax
   \else
   \ifx\diffarg\{%
   \let\DiffSpace\relax
   \fi\fi\fi\DiffSpace}

\newcommand{\widebar}{\overline}

\newcommand{\LR}{\Longleftrightarrow}

\renewcommand{\lim}[1]{\underset{#1}{\operatorname{lim}} \;}

\newcounter{abcd}[section]
\newcommand{\bitabc}{\vspace{-1.8ex}\begin{enumerate}
\renewcommand{\labelenumi}{\alph{abcd})}
\itemsep-1.4ex \partopsep-1.8ex}
\newcommand{\eitabc}{\end{enumerate}}

\newcommand{\bitp}{\vspace{-1ex}\begin{itemize}
\itemsep-0.6ex \partopsep-0.3ex}
\newcommand{\eitp}{\end{itemize}}

\newenvironment{fshaded}{\MakeFramed{\FrameRestore}}{\endMakeFramed}
\newenvironment{orangeumgebungs}[1][]{\definecolor{shadecolor}{rgb}{1,.8,.8}\definecolor{framecolor}{rgb}{1,0,0}\begin{fshaded}\begin{equation*}#1}{\end{equation*}\end{fshaded}}
\newcommand{\bous}{\begin{orangeumgebungs}}
\newcommand{\eous}{\end{orangeumgebungs}}

\newenvironment{blaueumgebungs}[1][]{\definecolor{shadecolor}{rgb}{.9,.9,1}%
\definecolor{framecolor}{rgb}{.1,.0,.7}%
\begin{fshaded}\begin{equation*}#1}{\end{equation*}\end{fshaded}}
\newcommand{\bbus}{\begin{blaueumgebungs}}
\newcommand{\ebus}{\end{blaueumgebungs}}

\newenvironment{blaueumgebunge}[1][]{\definecolor{shadecolor}{rgb}{.9,.9,1}%
\definecolor{framecolor}{rgb}{.1,.0,.7}%
\begin{fshaded}\begin{enumerate}#1}{\end{enumerate}\end{fshaded}}
\newcommand{\bbue}{\begin{blaueumgebunge}}
\newcommand{\ebue}{\end{blaueumgebunge}}

\newcommand{\beq}{\begin{equation}}
\newcommand{\eeq}{\end{equation}}
\newcommand{\beqs}{\begin{equation*}}
\newcommand{\eeqs}{\end{equation*}}

\newcommand{\bal}{\begin{align}}
\newcommand{\eal}{\end{align}}
\newcommand{\bals}{\begin{align*}}
\newcommand{\eals}{\end{align*}}

\newcommand{\mathbfat}{\begin{pmatrix}}
\newcommand{\emat}{\end{pmatrix}}

\newcommand{\bit}{\begin{itemize}}
\newcommand{\eit}{\end{itemize}}
\newcommand{\biit}{\begin{itemize}}
\newcommand{\eiit}{\end{itemize}}

\newcommand{\ben}{\begin{enumerate}}
\newcommand{\een}{\end{enumerate}}
\newcommand{\been}{\begin{enumerate}}
\newcommand{\eeen}{\end{enumerate}}

\newcommand{\bca}{\begin{cases}}
\newcommand{\eca}{\end{cases}}

\newcommand{\bpa}{\begin{parts}}
\newcommand{\epa}{\end{parts}}

\newcommand{\bspa}{\begin{subparts}}
\newcommand{\espa}{\end{subparts}}

 \newenvironment{NewSolution}
    {\SetTotalwidth\begin{solution}}
    {\end{solution}}
\newcommand{\bso}{\begin{NewSolution}}
\newcommand{\eso}{\end{NewSolution}}
		
\newtheorem*{remark}{Remark}

\newcommand{\blind}{0}

\newcommand*\samethanks[1][\value{footnote}]{\footnotemark[#1]}

\begin{document}

\def\spacingset#1{\renewcommand{\baselinestretch}%
  {#1}\small\normalsize} \spacingset{1}


\if0\blind
{
\title{\LARGE\bf Improving Estimation in Functional Linear Regression with Points of
  Impact: Insights into Google AdWords}
\author{
  Dominik Liebl\thanks{Institure of Finance and Statistics and Hausdorff Center for Mathematics, University of Bonn},
  Stefan Rameseder\thanks{Department of Econometrics, University of Regensburg},
  and Christoph Rust\samethanks
} 
\clearpage 
\maketitle
\thispagestyle{empty} 
} \fi

\if1\blind
{
  \renewcommand{\thefootnote}{\fnsymbol{footnote}}
  \vspace*{3cm}
  \begin{center}
   {\LARGE\bf Improving Estimation in Functional Linear Regression with Points of
  Impact: Insights into Google AdWords
  \end{center}
  \vspace*{3cm}
  \thispagestyle{empty}
} \fi

\bigskip

\begin{abstract}
The functional linear regression model with points of impact is a recent augmentation of the classical functional linear model with many practically important applications. In this work, however, we demonstrate that the existing data-driven procedure for estimating the parameters of this regression model can be very instable and inaccurate. The tendency to omit relevant points of impact is a particularly problematic aspect resulting in omitted-variable biases. We explain the theoretical reason for this problem and propose a new sequential estimation algorithm that leads to significantly improved estimation results. Our estimation algorithm is compared with the existing estimation procedure using an in-depth simulation study. The applicability is demonstrated using data from Google AdWords, today's most important platform for online advertisements. The \textsf{R}-package \texttt{FunRegPoI} and additional \textsf{R}-codes are provided in the online supplementary material. 
\end{abstract}

\noindent
{\it Keywords:} Functional data analysis, functional linear regression, points of impact, online advertising
\vfill

\newpage
\spacingset{1.5}

\section{Introduction}\label{sec:intro}
In many practical applications, one is interested in the relationship between a real-valued outcome variable $Y_i$ and a function-valued predictor $\{X_i(t);\;a\leq t\leq b\}$. In our motivating Google AdWords case study, for instance, we aim to explain the numbers of clicks $Y_i$ using impression trajectories $X_i(t)$, where $t$ denotes a certain day within the considered time interval $[a,b]$ of one year and $i=1,\dots,n$ indexes the cross section of keywords associated with the considered Google AdWords ad campaign.\footnote{Online ad campaigns use text corpora populations of relevant search keywords (for instance, \texttt{outdoor jacket}, \texttt{mountain boots}, etc., in the case of an outdoor equipment campaign) to identify potential customers by their Google searches (see Section \ref{sec:app} for more details).} The economic success of any ad campaign depends on product specific (time-global) seasonalities as well as on (time-local) events. The slowly varying seasonal component could be estimated using the function-valued slope parameter of the classical functional linear regression model \citep[see, e.g.,][]{HalHor2007}. The presence of time-local effects, however, harms such a simple estimation approach (see the right plot in Figure \ref{fig:dev} for notable examples). Therefore, we use the recent functional linear regression models with so-called Points of Impact (PoI) that allow us to identify and to control for time-local effects. 

Point of impact models are originally introduced by \cite{McKSen2010}, who argue that these models are better to interpret than the classical functional linear regression models. Indeed, several convincing real data applications are presented in the related work of \cite{LinMcK2009}. The method of \cite{KnePosSa2016} generalizes the original point of impact model by adding a classical functional linear regression component. While the original point of impact model captures only time-local effects, the augmented point of impact model of \cite{KnePosSa2016} allows also for time-global effects. In our paper we present a new and relevant case study where time-local as well as time-global effects are important for modeling the outcome.

As demonstrated in our simulation study, however, the finite sample performance of the estimation procedure proposed by \citet{KnePosSa2016} is very sensitive to the performance of the involved model selection. Therefore, we propose an adjusted sequential estimation algorithm that leads to significantly improved and more robust estimation results by using a refined model selection procedure.

The functional linear regression model with PoIs of \citet{KnePosSa2016} is related to several other works in the literature. Identifiability and estimation of points of impact was originally studied by \citet{McKSen2010}. The authors focus on a one-point of impact model without functional linear model component; however, the possibility of a partial model misspecification by an additional functional linear model component is also discussed theoretically.  \citet{FerHalVi2010} allow for multiple PoIs within a nonparametric model, but also do not consider a functional linear model component. \cite{MatKon2011} consider the extraction of local information within functional linear regressions using a LASSO-type approach, but do not estimate global components. \cite{TorBerCue2016} focus on a classification context, and \cite{FraYanMar2016} consider feature selection for functional data at a more general level. Our estimation algorithm uses the penalized smoothing splines estimator for functional linear regression models proposed by \cite{CraKneSa2009}. The related literature is extensive and the following examples are by no means exhaustive. \cite{CarCraKne2007} consider functional linear regression with errors-in-variables, \cite{CraKneSa2009} address optimality issues, \cite{GolBobCraCipCafRei2010} focus on penalized smoothing splines within a mixed model framework, and \cite{MarYoh2013} propose a robust version of the penalized smoothing splines estimator. Scalar-on-function regression models are successfully applied to solve important practical problems. \cite{Chi2012} proposes a functional regression model for predicting traffic flows. \cite{GolCraCafRei2012} introduce a penalized functional regression model to explore the relationship between cerebral white matter tracts in multiple-sclerosis patients. \cite{KoeZhuNanWan2014} consider regularized functional linear regression for brain image data. \cite{GelColNeeCrai2014} and \cite{GroKokSoj2017} propose functional regression models for incomplete curves. An overview article on methods for scalar-on-function regression is found in \cite{ReiColShaOgd2016}. Readers with a general interest in Functional Data Analysis (FDA) are referred to the textbooks of \cite{RamSil2005}, \cite{FerVie2006}, \cite{HorKok2012}, and \cite{HsiEub2015}. To the best of our knowledge, we are the first to use methods from FDA to analyze data from an online ad campaign; however, there are several contributions in FDA on related applications. \cite{RedDas2006} use a classical functional linear regression model to analyze online art auctions, \cite{LiuMue2008} analyze eBay auction prices using methods for sparse functional data, \citet{WanJanSh2008a} forecast eBay auction prices, \citet{WanJanSh2008b} develop a model for the price dynamics at eBay using differential equation models, and \cite{ZhaJanShm2010} consider real-time forecasting of eBay auctions using functional K-nearest neighbors.

The rest of the paper and our contributions are structured as follows. The next section (Section \ref{sec:met}) contains our methodological part.  In Section \ref{ssec:KPS}, we begin with a short presentation of the original procedure of \cite{KnePosSa2016}.  In Section \ref{ssec:estadj}, we introduce our three main proposals (1.~Sequential model selection and estimation, 2.~Smoothing splines estimator, and 3.~Standardizations) which we use to stabilize and improve the estimation procedure of \cite{KnePosSa2016}.  The implementation of our estimation algorithm is presented in Section \ref{ssec:PESES}. Section \ref{sec:sim} contains our simulation results, our case study on analyzing Google AdWords data is found in Section \ref{sec:app} and Section \ref{sec:con} concludes.  Appendix \ref{Appendix} presents further simulation results. The online supplement \cite{LieRamRus19} supporting this article contains the \textsf{R}-package \texttt{FunRegPoI} and the \textsf{R}-codes to reproduce our simulation study and the real data application.

\section{Methodology}\label{sec:met}
We formally consider the following functional linear regression model with PoIs introduced by \citet{KnePosSa2016}:
\begin{equation}
Y_i=\int_a^b \beta(t)X_i(t)dt + \sum_{s=1}^S \beta_s X_i(\tau_s) + \epsilon_i,\quad i = 1,\ldots,n. \label{kneip}
\end{equation}
Here, $(Y_1,X_1),\ldots,(Y_n,X_n)$ denote an i.i.d.~sample of scalar response variables $Y_i\in\mathbb{R}$ and random predictor functions $X_i\in L^2([a, b])$, where $\E{Y_i}=0$ and $\E{X_i(t)}=0$ for all $t\in[a,b]$. Without loss of generality, we set $[a,b]=[0,1]$. The i.i.d.~error term $\epsilon_i$ has mean zero, variance $\E{\epsilon_i^2}=\sigma^2_{\epsilon}<\infty$, and is independent of $X_i$. The assumption that $Y_i$ and $X_i$ have mean zero is only for notational simplicity; for the estimation, however, we will explicitly denote the centering of the data.

The function-valued slope parameter $\beta\in L^2([0,1])$ in Model \eqref{kneip} describes the time-global influences of $X_i$ on $Y_i$. The scalar-valued slope parameters $\beta_s\in\mathbb{R}$ take into account the time-local influences where the corresponding (unknown) time-points $\tau_s$ denote the locations of the PoIs. The estimation algorithm described below addresses the estimation of all unknown model parameters, namely, the global slope coefficient $\beta$, the local influences of the PoIs $\beta_1$,\ldots,$\beta_S$, and the set of PoI locations $\mathcal{T}=\{\tau_1$,\ldots,$\tau_S\}$.

In the following, we introduce our basic notation. The functions $X_i(t)$ are observed at $p$ equidistant grid points $t_1, \ldots, t_p$ with $t_j=(j-1)/(p-1)$. For non-equidistant designs, this can always be achieved by pre-smoothing the data. In $\mY = (Y_1, \ldots, Y_n)^\prime \in \R^n$, we collect all observations of the response variable $Y_i$, and in $\mX=(X_i(t_j))_{ij} \in\mathbb{R}^{n\times p}$, we collect all discretizations $X_i(t_j), i =1 , \ldots, n, j = 1, \ldots, p$. Furthermore, let $\mY^c$ and $\mX^c$ define the centered versions of $\mY$ and $\mX$, i.e., $\mY^c =(Y_1^c,\dots,Y_n^c)^\prime$, $\mX^c = (X_i^c(t_j))_{ij}$, where 
$Y_i^c=Y_i-\widebar{ \mY}$, 
$X_i^c(t_j)=X_i(t_j)-\widebar{ \mX}_j$,
$\widebar{\mY} = n^{-1}\sum_{i=1}^n Y_i$,
$\widebar{\mX}_{j} = n^{-1}\sum_{i=1}^n X_i(t_j)$. 

\subsection[]{The original procedure of \citet{KnePosSa2016}} \label{ssec:KPS}

In this section, we briefly describe the estimation and model selection procedures proposed in \citet{KnePosSa2016}.  Afterwards, in Section \ref{ssec:estadj},  we describe our adjustments to improve the original procedure and explain why these adjustments result in superior estimation performances.

\begin{figure}[!ht]
\centering
\includegraphics[width=\textwidth]{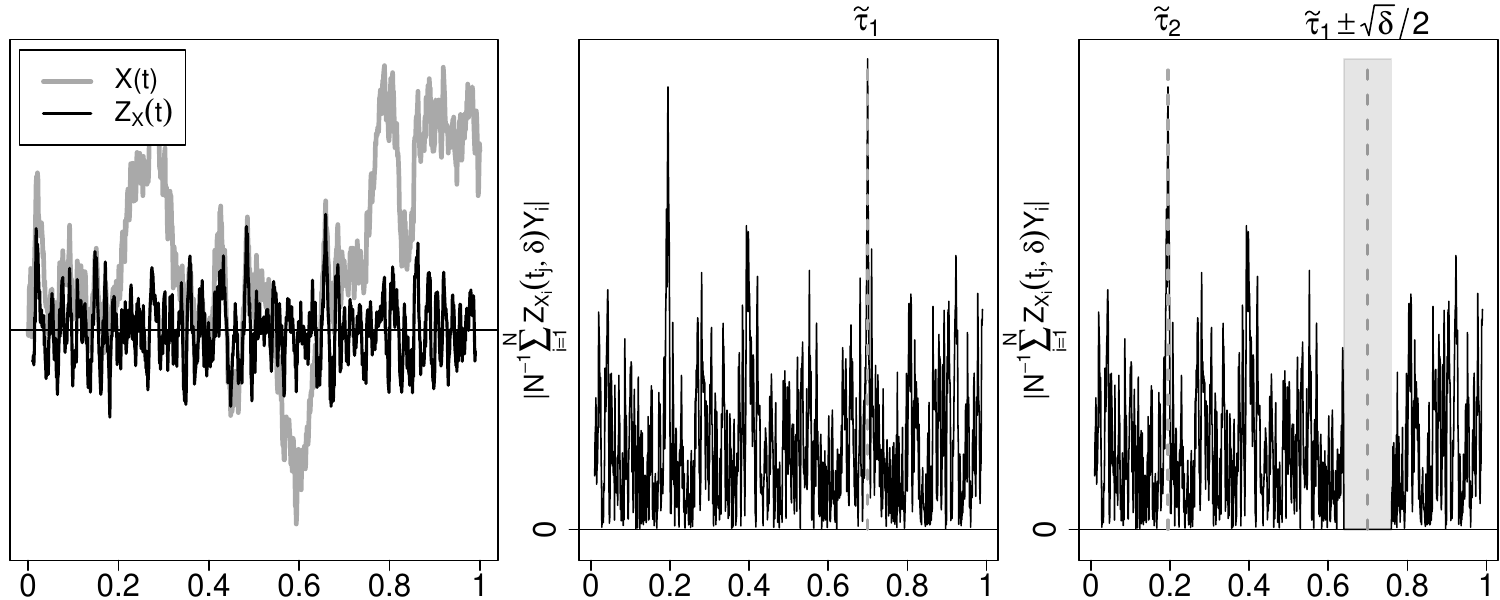}
\caption[searchPotPoi-Algorithm]{\textsc{Left:} Trajectories of $X_i(t_j)$ and $Z_{X_i}(t_j; \delta)$, with $\delta=0.01$. \textsc{Middle:} Trajectory of  $|n^{-1}\sum_{i=1}^n Z_{X_i}(t_j; \delta)Y_i|$, with first choice $\widetilde{\tau}_1$. \textsc{Right:} Visualization of the second iteration of the \texttt{searchPotPoI}-Algorithm.}\label{fig:decorrelation}
\end{figure}

To estimate the potential PoIs $\widetilde{\tau}_s, s = 1, \ldots, \widetilde{S}$, \citet{KnePosSa2016} propose a local maxima search (over $t_j$) based on the sample version $|n^{-1}\sum_{i=1}^n Z_{X_i}(t_j; \delta)Y_i|$ of the cross-moment $|\E{Z_{X_i}(t; \delta) Y_i}|$, where $Z_{X_i}(t; \delta) = X_i(t) - (X_i(t-\delta) + X_i(t+\delta))/2$ is the central second-order difference quotient of $X_i(t)$ with $\delta>0$.  The statistic $Z_{X_i}(t; \delta)$ acts as a filter on $X_i(t)$ that uncovers the local-specific variance component of the process $X_i(t)$;  see the left plot in Figure \ref{fig:decorrelation}.

The existence of a local-specific variance component in $X_i$ is crucial for the estimation procedure of \cite{KnePosSa2016} and allows to show the identifiability of the points of impact and of the model parameters model parameters \citep[see][Theorem 1]{KnePosSa2016}.  Processes that have a local-specific variance component are typically rough stochastic processes (for instance, Brownian motions, Ornstein-Uhlenbeck processes, etc.), i.e., processes with covariance functions that are sufficiently non-smooth at the diagonal \citep[see][Theorem 3]{KnePosSa2016}. \cite{KnePosSa2016} use a parameter $0<\kappa<2$ to quantify the smoothness of the covariance function at the diagonal and propose an estimator $\widehat\kappa$ to decide in practice, whether the covariance function is sufficiently non-smooth at the diagonal. The reader is referred to Section \ref{sec:app} for an application of this procedure.

\begin{algorithm}[!tb]
\caption{Search Potential Points of Impact Algorithm}\label{alg:potPoI} \
\begin{algorithmic}[1]
\Procedure{searchPotPoI}{\hspace{1mm}$\delta \in \calD = (0, \delta_{\text{max}}], \hspace{1mm} \mX = \mX^{c}, \hspace{1mm} \mY=\mY^{c}$} 
\State Given $\delta$, define the index $k_{\delta} \in \N$ such that $1 \leq k_\delta < (p-1)/2 \LR \delta \approx k_\delta /(p-1)$.
\State Restrict the set of possible grid indices, i.e., define $\calJ_{0,\delta} = \lbrace k_\delta+1, \ldots , p-k_\delta \rbrace$.
\State For each index $j \in \calJ_{0, \delta}$, calculate $Z_{X_i}(t_j; \delta) = X_i(t_j) - \frac{1}{2}(X_i(t_j-\delta) + X_i(t_j+\delta)).$ 
\While{$\calJ_{s, \delta} \neq \emptyset$, iterate over $s = 1, 2, 3, \ldots,$ and }
\State Determine the index $j_s \in \calJ_{s-1, \delta}$ of the empirical maximum of $Z_{X}(t; \delta) Y$, i.e., 
\vspace{-3mm}
\beqs
j_s = \underset{j \in \calJ_{s-1, \delta}}{\text{argmax}}\left|\frac{1}{n} \sum_{i=1}^n Z_{X_i}(t_j; \delta) Y_i\right|.
\eeqs
\State Define the $s$-th potential impact point $\widetilde{\tau}_s=t_{j_{s}}$ as grid point at index $j_s$.
\State Eliminate all points in an environment of size $\sqrt{\delta}$ around $\widetilde{\tau}_{s}$, i.e., define 
\vspace{-3mm}
\beqs 
\calJ_{s,\delta} = \lbrace j \in \calJ_{s-1,\delta}\mid |t_j-\widetilde{\tau}_{s}| \geq \sqrt{\delta}/2\rbrace.
\eeqs
\EndWhile
\State \textbf{return} $\widetilde{\calT}=\lbrace\widetilde{\tau}_1, \ldots, \widetilde{\tau}_{\widetilde{S}}\rbrace$ \ 
\EndProcedure
\end{algorithmic}
\end{algorithm} 

The estimation procedure proposed by \citet{KnePosSa2016} to detect potential PoIs is formally described in Algorithm \ref{alg:potPoI}.  In each iteration, $s=1,2,\dots$, one PoI is selected by the global maximum of the trajectory of $|n^{-1}\sum_{i=1}^n Z_{X_i}(t_j; \delta)Y_i|$ over $j\in\calJ_{s-1,\delta}$, where $\calJ_{s-1,\delta}\subset\{1,\dots,p\}$ denotes an index set defined in Algorithm \ref{alg:potPoI} (see lines 3 and 8).  Once a PoI is selected, the algorithm eliminates the grid points within a $\sqrt{\delta}/2$-neighborhood around the selected PoI (see line 8 of Algorithm \ref{alg:potPoI}). The algorithm terminates when $\calJ_{s,\delta}$ is the empty set. The elimination step in line 8 is necessary for providing a consistent estimation procedure.

The selection of the first PoI is shown in the middle plot of Figure \ref{fig:decorrelation}. The first elimination step is shown in the right plot of Figure \ref{fig:decorrelation}, where the second PoI, $\widetilde{\tau}_2$, is determined by the global maximum of the remaining parts of the trajectory of $|n^{-1}\sum_{i=1}^n Z_{X_i}(t_j;\delta)Y_i|$ over $j\in\calJ_{1,\delta}$.

To estimate the model coefficients for given PoIs $\widehat{\tau}_s$, \citet{KnePosSa2016} propose an FPCA-based estimation procedure using the approximate model $Y_i \approx \allowbreak \int_0^1\beta_{K}(t) X_{i,K}(t)dt +\allowbreak \sum_{s=1}^{\widehat{S}}\beta_s X_i(\widehat{\tau}_s)+\epsilon_i$, where $\beta_{K}(t)\approx\beta(t)$ and $X_{i,K}(t)\approx X_i(t)$ are $K$-dimensional approximations based on the first $K$ eigenfunctions of the empirical covariance operator of $X_i$ \citep[see][Eq.~(6.1)]{KnePosSa2016}.
Besides the smoothing parameter, $K$, one needs to choose a good value of the tuning parameter $\delta$ and a subset $\widehat{\calT}\subseteq\widetilde{\calT}$ of the set of potential PoIs $\widetilde{\calT}$ from Algorithm \ref{alg:potPoI}. For selecting $\widehat{\calT}\subseteq\widetilde{\calT}$, \citet{KnePosSa2016} propose an asymptotic cut-off approach and data-driven Bayesian Information Criterion (BIC)-based approach.  In this paper, we focus on the data-driven BIC-based approach as this approach performs clearly better than the asymptotic approach \citep[see Table 1 in][]{KnePosSa2016}.  Moreover, the asymptotic cut-off parameter is hardly applicable in practice as it depends on a generally unknown constant $A>\sqrt{2}$ \citep[see Table 4 in][]{KnePosSa2016}. 

\citet{KnePosSa2016} propose an infeasible version and a more general feasible version of their data-driven BIC-based procedure to select $K$, $\delta$, and $\widehat{\calT}\subseteq\widetilde{\calT}$.  The infeasible strategy is used in their simulation study where the authors perform a $\text{BIC}$-based selection of $K$ and $\widehat{\calT}$, and set $\delta=1/\sqrt{n}$.  This naive parametrization of $\delta$ is appropriate in their simulation study, but can be arbitrarily bad in practice.  The more general and strategy is used in the application section of \cite{KnePosSa2016} where the authors optimize the $\text{BIC}(K,\widehat{\calT},\delta)$ \emph{simultaneously} over $K$, subsets $\widehat{\calT}\subset\widetilde{\calT}$, and a fine grid of $\delta\in(0,\delta_{\text{max}}]$.  In this paper, we only focus on the latter general model selection strategy since this is the practically most relevant strategy proposed in \cite{KnePosSa2016} which is not based on unknown constants or naive choices of tuning parameters.



\subsection[]{Improving the procedure of \citet{KnePosSa2016}} \label{ssec:estadj}  
In this section, we explain our three main proposals to improve the estimation procedure of \citet{KnePosSa2016}: 1.~Sequential model selection and estimation, 2.~Smoothing splines estimator, and 3.~Standardizations.  Afterwards, in Section \ref{ssec:PESES},  we describe the implementation our estimation algorithm which builds upon these proposals.

\begin{figure}[htb]
\centering
\includegraphics[width=\textwidth]{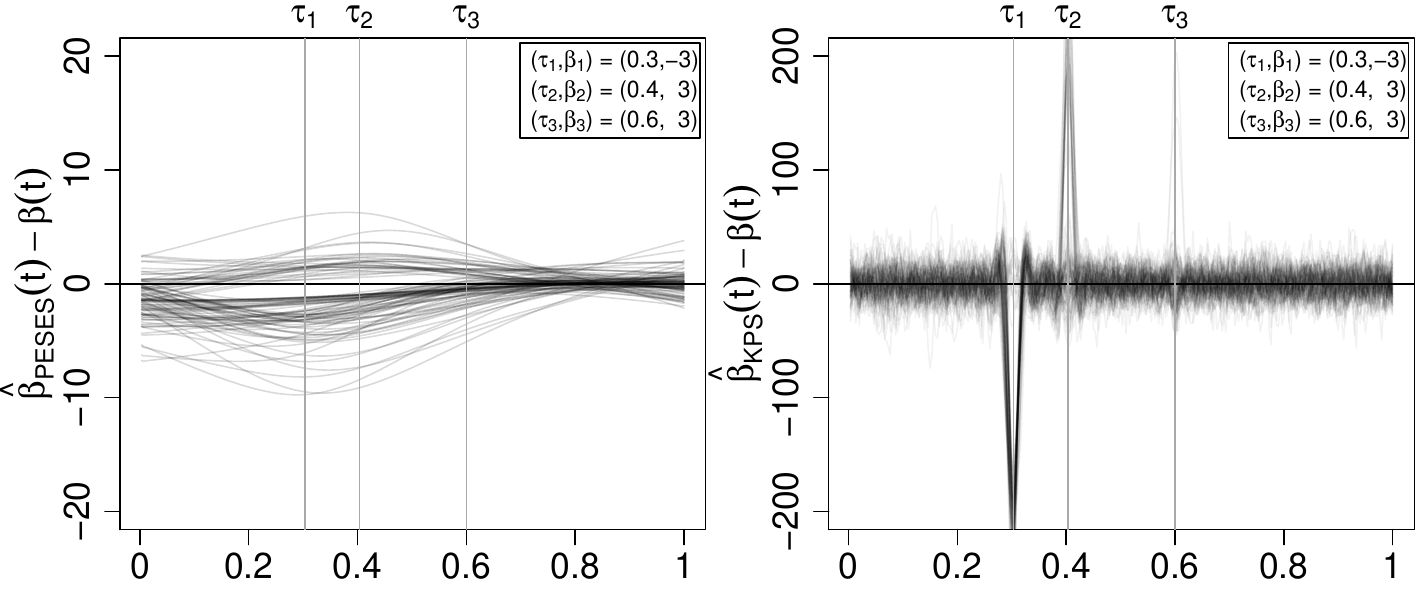}
\caption[]{Pointwise deviations $\widehat{\beta}(t)-\beta(t)$ of the 10\% largest $L^2$ distances $\int_0^1(\widehat{\beta}(t)-\beta(t))^2dt$ for the Complicated DGP. Note that the scales of the two y-axes differ by a factor of 10.}\label{fig:dev}
\end{figure}

\noindent\textbf{1.~Sequential model selection and estimation.} Estimating the model parameters in Model \eqref{kneip} bears the substantial risk of an omitted-variable-bias since not incorporating the (unknown) true PoI locations $\tau_s$ can result in a heavily biased estimator $\widehat{\beta}(t)$ (see the right plot in Figure \ref{fig:dev} for noteworthy examples). This is a critical issue in practice, and our simulation results show that the original estimation procedure of \citet{KnePosSa2016} may suffer severely from such biases.

The underlying problem is that the selection of the number $\widehat{S}$ of PoIs and their locations $\widetilde{\tau}_1,\ldots,\widetilde{\tau}_{\widehat{S}}$ and the selection of the smoothing parameter, $K$, for estimating $\beta(\cdot)$ are two ambiguous selection problems.  It is easy to trade model complexities between the empirical PoI model component and the empirical functional model component without affecting the model fit.  This results in a quite delicate model selection problem which generally leads to instable estimates when trying to solve both selection problems simultaneously as suggested in \citet{KnePosSa2016}.

Let us explain the reason for this instability by considering the following two extreme situations---both approximating the regression Model~\eqref{kneip}:
\begin{itemize}[leftmargin=.35cm, labelwidth=!, labelindent=0pt]  
\item Let $K\gg 0$ and $\widehat{S}=0$. For very large $K$ the estimator $\widehat{\beta}_K(t)$ is flexible enough, such that  
\begin{equation*}
\int_0^1\widehat{\beta}_K(t)X_{i,K}(t)dt\approx\int_0^1 \beta(t)X_i(t) dt + \sum_{s=1}^S\beta_s X_i(\tau_s).
\end{equation*} 
In this case, $\widehat{\beta}_K(t)$ approximates $\beta(t)$, except at the points of impact locations $t=\tau_s$, where $\widehat{\beta}_K(t)$ approximates $\beta_s X_i(\tau_s)$, i.e., where $\int_{\tau_s-h}^{\tau_s+h}\widehat{\beta}_K(t)X_{i,K}(t)dt\approx\beta_s X_i(\tau_s)$ with, e.g., $h=0.01$ (see the right plot in Figure \ref{fig:dev} for examples of such estimates $\widehat{\beta}_K(t)$). 

\smallskip

\item Let $K=0$ and $\widehat{S}\gg 0$. A large set of point of impact candidates $X_i(\widehat{\tau}_1),\dots,X_i(\widehat{\tau}_{\widehat{S}})$ leads to a very flexible linear model, such that 
\begin{equation*}
\sum_{s=1}^{\widehat{S}}\widehat{\beta}_s X_i(\widehat{\tau}_s)\approx \int_0^1 \beta(t)X_i(t) dt + \sum_{s=1}^S \beta_s X_i(\tau_s).
\end{equation*}
In this case, $\sum_{s=1}^{\widehat{S}}\widehat{\beta}_s X_i(\widehat{\tau}_s)$ acts like a Riemann sum for approximating $\int_0^1 \beta(t)X_i(t)dt$, except for the $\widehat{\beta}_s$-values at $\widehat{\tau}_s\approx\tau_s$, where $\widehat{\beta}_sX_i(\widehat{\tau}_s)\approx\beta_s X_i(\tau_s)$.
\end{itemize}

These two extreme situations demonstrate that there is a certain ambiguity between the model selection parameters $K$ and $\widehat{S}=|\widehat{\calT}(\delta)|$ that allows for shifting the model-complexities between the integral-part and the PoI-part of the empirical model. This ambiguity generally leads to unstable model selections when optimizing $\text{BIC}(K,\widehat{\calT},\delta)$ \emph{simultaneously} over $K$, subsets $\widehat{\calT}\subset\widetilde{\calT}$, and $\delta$---as proposed in \citet{KnePosSa2016}.  As a consequence, one gets instable estimates of $\beta(\cdot)$ caused by omitted-variable biases in $\widehat\beta(\cdot)$, as shown in the right plot in Figure \ref{fig:dev}.

To stabilize the model selection procedure we propose a sequential selection and estimation procedure (see Section \ref{ssec:PESES}).  In the first (``Pre-select'') step, our procedure pre-selects all potential points of impact $\widetilde{\calT}=\{\widetilde{\tau}_1,\dots,\widetilde{\tau}_{\widetilde{S}}\}$ while ignoring the estimation of the functional parameter $\beta(\cdot)$.  In the second (``Estimate'') step, our procedure estimates the model parameters, $\beta(\cdot)$ and $\beta_s$, given the pre-selected points of impact. 

The theoretical justification for this sequential approach is given by the following result which holds under the assumptions of \cite{KnePosSa2016} and implies that the points of impact can be estimated consistently without knowledge (or pre-estimation) of the slope function $\beta(\cdot)$ \citep[see Lemmas 3 and 4 in the supplemenary paper of][]{KnePosSa2016}:
\begin{equation}\label{eq:asympIdent}
\begin{array}{ll}
|\mathbb{E}(Z_{\delta,i}(t)Y_i)| = \beta_r c(\tau_s) \delta^\kappa+o(\delta^\kappa)&\text{if \;$t_j=\tau_s$ for some $s=1,\dots,S$}\\
|\mathbb{E}(Z_{\delta,i}(t)Y_i)| = O(\delta^2) &\text{if \;$t\not\in\{\tau_1,\dots,\tau_S\}$}
\end{array}
\end{equation}
as $\delta\to 0$, where $0<\kappa<2$ and $0<c(\tau_r)<\infty$ are constants specific to the considered process $X_i$. 

That is, the trajectory of $|\mathbb{E}(Z_{\delta,i}(t_j))Y_i)|$, $j=1,\dots,p$, will have peaks at grid points $t_j\approx \tau_r$, even without knowledge (or pre-estimation) of the slope function $\beta(\cdot)$. Consequently, Step 1 (``Pre-select'') of our algorithm (Section \ref{ssec:PESES}) leads to a consistent point of impact selection if, for instance, $\delta^\kappa\sim n^{-1}$, since $|\mathbb{E}(Z_{\delta,i}(t_j))Y_i)|$ can be consistently estimated using its empirical counterpart $|n^{-1}\sum_{i=1}^nZ_{\delta,i}(t_j))Y_i|$ for all $j=1,\dots,p$ as $n\to\infty$.

Using the consistently pre-selected points of impact in Step 2 (``Estimate'') leads to a more stable estimation of the model parameters, $\beta(\cdot)$ and $\beta_s$, as it avoids a simultaneous selection of the PoIs and the smoothing parameters.  The further Steps 3-5 (see the overview in Section \ref{PESES}) of our selection and estimation algorithm in Section contain repetitions of the selection and estimation steps (Step 1 and Step 2). These repetitions are asymptotically irrelevant, but further improve the estimation results in practice (see Section \ref{sec:sim}).  

\noindent\textbf{2.~Smoothing splines estimator.} Deviating from \citet{KnePosSa2016}, we propose using a penalized smoothing splines estimator.  The FPCA-based estimator, proposed by \citet{KnePosSa2016}, is optimal only under the restrictive assumption of a structural link between the functional regression parameter, $\beta(\cdot)$, and the functional regressor, $X$ \citep[see Assumptions (3.1-3.3) in][]{HalHor2007}.  However, this link does not necessarily hold in applications and also cannot be tested in practice.\footnote{Remember that the FPCA-basis, based on the eigendecomposition of the covariance operator of $X$, is the optimal empirical basis to approximate $X$, but generally not the optimal basis to approximate $\beta(\cdot)$.} Therefore, we propose using the penalized smoothing splines estimator of \cite{CraKneSa2009}.  While this estimator achieves minimax optimal rates under similar structural link assumptions (see Crambes et al.~, 2009), it is also known to perform well if these structural assumptions do not hold since the spline basis system has some very general approximation properties (see, for instance, \citeauthor{Boor2001}, \citeyear{Boor2001}, and \citeauthor{CraKneSa2009}, \citeyear{CraKneSa2009}). 

\noindent\textbf{3.~Standardizations.} The standardization of the curves in Step 1 (``Pre-select'') and Step 3 (``Sub-select'') of our algorithm scales the trajectories of the process $Z_\delta$ by the inverse of the pointwise standard deviation of the process $X$.   From an asymptotic perspective, this is irrelevant, since this scaling only leads to different constants $c(\tau_r)$ in \eqref{eq:asympIdent}.  However, standardization of the data is a typical pre-processing step in model selection problems leading to more homogenous signals which further improves the selection results in practice (see Section \ref{sec:sim}).



\subsection{The PES-ES estimation algorithm}\label{ssec:PESES}

Our estimation algorithm is built up from the following three Pre-select, Estimate, and Sub-select (PES) steps:  

\begin{description}
\item[1.~Pre-select:] Pre-select potential PoIs $\widetilde{\mathcal{T}}=\{\widetilde{\tau}_1,\ldots,\widetilde{\tau}_{\tilde{S}}\}$. (See Section \ref{ssec:presel})
\item[2.~Estimate:] Estimate the function- and scalar-valued slope parameters \\
$\beta,\beta_1,\dots,\beta_{\tilde{S}}$ given the set of potential PoIs $\widetilde{\mathcal{T}}$. (See Section \ref{ssec:estbeta})
\item[3.~Sub-select:] Sub-select PoIs from the set of potential PoIs $\widetilde{\mathcal{T}}$. (See Section \ref{ssec:mod})
\end{description}

Typically, the estimation step (Step 2) leads to inefficient estimators $\widehat\beta(\cdot)$, but avoids omitted-variable biases. Inefficient, because $\widetilde{\calT}$ tends to contain many redundant PoI locations ($\widetilde{S}>S$), which reduces the number of degrees of freedom. We reduce the risk of omitted-variable biases, because the large set of potential PoIs $\widetilde{\mathcal{T}}$ has a high likelihood of containing the true PoI locations (as explained in more detail in Section \ref{ssec:estadj}). Our final PES-ES algorithm, described in Section \ref{PESES}, uses a repetition of the latter two Estimate-Sub-select (ES) steps, which can result in a further improvement of the estimation results (see Section \ref{sec:sim}).

\subsubsection{Pre-Select PoIs}\label{ssec:presel}

To select potential PoIs, we use Algorithm \ref{alg:potPoI} with the difference that instead of using centered observations of the functions $\mX^c$, we use the pointwise standardized curves $\mX^{st}$ as input of the algorithm, where $X_i^{st}(t_j)=X_i^c(t_j)/\text{sd}( \mX_j)$ and $\text{sd}(\mX_j)=(n^{-1}\sum_{i=1}^n(X_i(t_j)-\widebar{\mX}_{j})^2)^{1/2}$. As described in Section \ref{ssec:estadj}, this is irrelevant from an asymptotic point of view, but typically stabilizes and improves the PoI selection in practice. 

\subsubsection{Estimate Slope Parameters}\label{ssec:estbeta}

To estimate the slope parameters---given the pre-selected PoIs $\widetilde{\calT}$---we adapt the penalized smoothing splines estimator proposed by \citet{CraKneSa2009} in order to incorporate PoIs. Let us initially recap the situation of Model \eqref{kneip} without PoIs ($S=0$, $\calT=\emptyset$), as considered by \citet{CraKneSa2009}. Their estimator of $\beta(\cdot)$, evaluated at the grid points $t_1,\dots,t_p$, is given by
\begin{equation}
\big(\widehat{\beta}^\rho(t_1), \ldots, \widehat{\beta}^\rho(t_p)\big)= \frac{1}{n} \left(\frac{1}{np} \mX^{c\prime}\mX^c + \rho \mA \right)^{-1} \mX^{c\prime}\mY^c,\label{betahat_CKS}
\end{equation}
where the penalty matrix $\mA = \mP + p\mA^\star$ is composed of a non-classical projection matrix $\mP$ and a classical regularization matrix $\mA^\star$. The non-classical $p \times p$ projection matrix $\mP= \mW ( \mW^\prime \mW)^{-1} \mW^\prime$, with $\mW = (t_j^l)_{j, l} \in \R^{p \times m}$ is introduced by \citet{CraKneSa2009} in order to guarantee uniqueness of their estimator, where $t_j^l$ denotes the $l$th power of the grid point $t_j$ with $j=1,\dots,p$ and $l=0,\dots,m-1$.
Following the usual convention, we set $m=2$, which results in the classical choice of \emph{cubic} splines. The classical  $p\times p$ regularization matrix $\mA^\star$ is defined as
\begin{equation*}
\mA^\star = \mB (\mB^\prime\mB)^{-1} \left(\int_0^1 \vb^{(2)}(t) \vb^{(2)}(t)^\prime dt \right)  (\mB^\prime\mB)^{-1} \mB^\prime,
\end{equation*}
where $\vb(t) = (b_1(t), \ldots, b_p(t))^\prime$ are natural cubic spline basis functions, $\vb^{(2)}(t)$ denotes their second derivatives, and $\mB$ is a $p \times p$ matrix with elements $b_i(t_j), i, j = 1, \ldots, p$. For the implementation of the natural cubic spline basis functions, we use the \texttt{ns}-function contained in the \textsf{R}-package \texttt{splines}.

In order to incorporate the pre-selected PoIs, we need to extend the matrices $\mX^c$ and $\mA$. The extended data matrix is given by $\mX_{\widetilde{\calT}}^c = (\mX^c, p\mX^c(\widetilde{\tau}_1),\dots,\allowbreak p\mX^c(\widetilde{\tau}_{\widetilde{S}}))\in \R^{n\times (p+\widetilde{S})}$, where $\mX^c(\widetilde{\tau}_s)=(X_1^c(\widetilde{\tau}_s), \ldots, X_n^c(\widetilde{\tau}_s))^\prime \in \R^n$. The extended penalty matrix is given by
\begin{equation*}
\mA_{\widetilde{\calT}}=\begin{pmatrix} \mA & 0 \\ 0 & 0\end{pmatrix} \in \R^{(p+\widetilde{S}) \times (p+\widetilde{S})},
\end{equation*}
where all entries with respect to the PoIs are zero \citep[see][for an equivalent extension of the penalty matrix]{GolBobCraCipCafRei2010}. The augmented estimator of $\beta(t_1),\dots,\beta(t_p)$ and $\beta_1,\dots,\beta_S$,
\begin{equation}
\widehat{\beta}^\rho_{\widetilde{\calT}}=\big(\widehat{\beta}^\rho_{\widetilde{\calT}}(t_1), \ldots, \widehat{\beta}^\rho_{\widetilde{\calT}}(t_p), \widehat{\beta}_{\widetilde{\calT},1}^\rho, \ldots, \widehat{\beta}_{\widetilde{\calT},\widetilde{S}}^\rho\big) = \frac{1}{n} \left(\frac{1}{np} \mX_{\widetilde{\calT}}^{c\prime}\mX^c_{\widetilde{\calT}} + \rho \mA_{\widetilde{\calT}} \right)^{-1} \mX_{\widetilde{\calT}}^{c\prime}\mY^c,\label{betahat_0}
\end{equation}
depends on the included set of PoIs $\widetilde{\calT}$ and on the smoothing parameter $\rho$. In order to determine an optimal smoothing parameter, we use the following Generalized Cross-Validation (GCV) criterion, as proposed by \citet{CraKneSa2009}:
\begin{equation}
\text{GCV}(\rho) = \frac{\frac{1}{n}\text{RSS}(\widehat{\vbeta}_{\widetilde{\calT}}^\rho)}{\left(1-\frac{1}{n}\text{Tr}(\mH^c_{\rho,\widetilde{\calT}})\right)^2}.\label{gcv}
\end{equation}
Here, the Residual Sum of Squares (RSS) is defined as $\text{RSS}(\widehat{\vbeta}_{\widetilde{\calT}}^\rho)=\vert\vert\mY^c - \mH^c_{\rho,\widetilde{\calT}}  \mY^c \vert\vert^2$, where $||.||$ denotes the Euclidean norm, and the smoother matrix $\mH^c_{\rho,\widetilde{\calT}}$ is defined as $\mH^c_{\rho,\widetilde{\calT}} = (np)^{-1} \mX^c_{\widetilde{\calT}} ((np)^{-1}\mX_{\widetilde{\calT}}^{c\prime} \mX^c_{\widetilde{\calT}}+ \rho \mA_{\widetilde{\calT}})^{-1} \mX_{\widetilde{\calT}}^{c\prime}$. Our final estimator for the slope parameters is given by the GCV-optimized version of \eqref{betahat_0},
\begin{equation}
\widehat{\beta}_{\widetilde{\calT}}=\big(\widehat{\beta}_{\widetilde{\calT}}(t),\widehat{\beta}_{\widetilde{\calT},1}, \ldots, \widehat{\beta}_{\widetilde{\calT},\widetilde{S}}\big)=
\big(\widehat{\beta}^{\rho_{\text{GCV}}}_{\widetilde{\calT}}(t), \widehat{\beta}_{\widetilde{\calT},1}^{\rho_{\text{GCV}}}, \ldots, \widehat{\beta}_{\widetilde{\calT},\widetilde{S}}^{\rho_{\text{GCV}}}\big),\quad t\in\{t_1,\dots,t_p\},\label{betahat}
\end{equation}
where $\rho_{\text{GCV}}=\text{argmin}_{\rho\in(0,\rho_{\text{max}}]}\text{GCV}(\rho)$.

\subsubsection{Sub-Select PoIs} \label{ssec:mod}
This part of our estimation algorithm is aimed at selecting the true PoIs from the pre-selected set of potential PoIs $\widetilde{\calT}=\widetilde{\calT}(\delta)$ given the estimate $\widehat{\vbeta}_{\widetilde{\calT}}$ in \eqref{betahat}. This sub-selection is performed by minimizing the following BIC over subsets $\calR\subseteq\widetilde{\calT}(\delta)$:
\begin{equation*}
\widehat{\calT}=\underset{\calR\subseteq\widetilde{\calT}(\delta)}{\text{argmin}}\;\text{BIC}(\calR),\quad\text{where}
\end{equation*}
\begin{equation}
\text{BIC}(\calR) = n\log\left(\frac{\text{RSS}(\calR)}{n} \right) + \log(n) \cdot S_{\calR},\quad\text{with}\quad S_\calR=|\mathcal{R}|.\label{bic}
\end{equation}
Here, $\text{RSS}(\calR)$ is made up of the residuals from regressing the $\widehat{\vbeta}_{\widetilde{\calT}}$-neutralized $Y_{i, \widehat{\vbeta}_{\widetilde{\calT}}}= Y^c_i - \int_0^1 \widehat{\vbeta}_{\widetilde{\calT}}(t) X_i^c(t) dt$ onto $X^{st}_i(\widetilde{\tau}_{s}),\dots,X^{st}_i(\widetilde{\tau}_{S_\calR})$, with $\{\widetilde{\tau}_{1},\dots,\widetilde{\tau}_{S_\calR}\}=\calR$, where $\widehat{\vbeta}_{\widetilde{\calT}}(t)$ is the estimate of $\beta(\cdot)$ defined as the first element in the vector of estimates \eqref{betahat}. 

For optimizing $\text{BIC}(\calR)$ over $\calR\subseteq\widetilde{\calT}(\delta)$, we use a directed search strategy taking into account the information content in $\widetilde{\calT}=\{\widetilde{\tau}_{1},\dots,\widetilde{\tau}_{\widetilde{S}}\}$. By construction, the order of the PoI locations $\widetilde{\tau}_{1},\dots,\widetilde{\tau}_{\widetilde{S}}$ reflects a decreasing signal-to-noise ratio and, therefore, a decreasing quality of the estimates. This suggests minimizing $\text{BIC}(\calR)$ using a directed search strategy where $\text{BIC}(\calR)$ is evaluated consecutively at the sets
$
\calR=\{\widetilde{\tau}_{1}\},\;
\calR=\{\widetilde{\tau}_{1},\widetilde{\tau}_{2}\},\dots,\calR=\{\widetilde{\tau}_{1},\dots,\widetilde{\tau}_{\widetilde{S}}\}.
$

\subsubsection{The full PES-ES estimator}\label{PESES}
Our estimation algorithm, PES-ES, consists of the above described Pre-select-Estimate-Sub-select (PES) steps and uses a repetition of the latter two Estimate-Sub-select (ES) steps:
\begin{center}
\begin{tabular}{lll}
1.~\textbf{P}re-Select   & $\widetilde{\calT}=\widetilde{\calT}(\delta)$  & (Section \ref{ssec:presel})\\
2.~\textbf{E}stimate     & $\widehat{\beta}_{\widetilde{\calT}}$          & (Section \ref{ssec:estbeta})\\
3.~\textbf{S}ub-Select   & $\widehat{\calT}\subseteq\widetilde{\calT}$    & (Section \ref{ssec:mod})\\
4.~re\textbf{E}stimate   & $\widehat{\beta}_{\widehat{\calT}}$            & (Section \ref{ssec:estbeta}, with $\widetilde{\calT}$ replaced by $\widehat{\calT}$)\\
5.~re\textbf{S}ub-Select & $\widehat{\calT}_{\text{re}}\subseteq\widehat{\calT}$ & (Section \ref{ssec:mod}, with $\widetilde{\calT}$ replaced by $\widehat{\calT}$)\\
\end{tabular}
\end{center}

Note that the entire PES-ES algorithm depends on the initially pre-selected set of potential PoIs $\widetilde{\calT}(\delta)$ and, therefore, on the choice of $\delta$. In the following, we write $\widehat{\calT}_{\text{re}}(\delta)$ in order to emphasize this entire dependency on $\delta$. We follow \citet{KnePosSa2016} and determine an optimal $\delta$ by minimizing the BIC. For each $\delta$-value on a fine grid in $(0,\delta_{\text{max}}]$, we run the entire PES-ES algorithm and select the optimal $\delta$ by,
\begin{equation*}
\delta_{\text{BIC}}=\underset{\delta\in(0,\delta_{\text{max}}]}{\text{argmin}}\;\text{BIC}(\delta),\quad\text{with}
\end{equation*}
\begin{equation}
\text{BIC}(\delta) = n\log\left(\frac{\text{RSS}(\widehat{\calT}_{\text{re}}(\delta))}{n} \right) + \log(n) \cdot \text{edf}(\widehat{\calT}_{\text{re}}(\delta)),\label{bicSim}
\end{equation}
where $\text{RSS}(\widehat{\calT}_{\text{re}}(\delta))=\vert\vert\mY^c - \mH^c_{\rho_{\text{GCV}},\widehat{\calT}_{\text{re}}(\delta)}  \mY^c\vert\vert^2$ with smoother matrix $\mH_{\rho_{\text{GCV}},\widehat{\calT}_{\text{re}}(\delta)}^c$ defined as $\mH_{\rho_{\text{GCV}},\widehat{\calT}_{\text{re}}(\delta)}^c = (np)^{-1} \mX_{\widehat{\calT}_{\text{re}}(\delta)}^c((np)^{-1}\mX^{c\prime}_{\widehat{\calT}_{\text{re}}(\delta)}\mX_{\widehat{\calT}_{\text{re}}(\delta)}^c+ \rho_{\text{GCV}} \mA)^{-1}\mX^{c\prime}_{\widehat{\calT}_{\text{re}}(\delta)}$ and effective degrees of freedom $\text{edf}(\widehat{\calT}_{\text{re}}(\delta)) = \text{Tr}(\mH^{c\prime}_{\rho_{\text{GCV}},\widehat{\calT}_{\text{re}}(\delta)}\mH_{\rho_{\text{GCV}},\widehat{\calT}_{\text{re}}(\delta)}^c)$; see \cite{HasTreTib1990}, Ch.~3.5 for an overview of possible definitions of $\text{edf}$.

\section{Simulations}\label{sec:sim}
In the following simulation study, we assess the finite sample properties of our PES-ES algorithm. The original estimation procedure proposed by \citet*{KnePosSa2016}, abbreviated as KPS, serves as our main benchmark, and its implementation is described in Section \ref{ssec:KPS}. The smoothing splines estimator \eqref{betahat_CKS} by \citet*{CraKneSa2009}, abbreviated hereafter as CKS, serves as a challenging benchmark for our NoPoI data generating process (i.e., a functional linear regression model \emph{without} points of impact). Section \ref{ssec:DGPandResults} introduces the considered data generating processes and presents our simulation results. 

We aim to provide an in-depth assessment of our PES-ES estimation algorithm. Therefore, in order to assess the improvements that are due to the final ES (Estimation and Subselection) step, we compare the PES-ES results with those of the reduced PES estimation algorithm without the final ES step. We also show that an additional second repetition of the ES step (PES-2ES) does not lead to a significant improvement of our estimation algorithm.

\citet{KnePosSa2016} arbitrarily set $K_{\text{max}}=6$, which is, however, too small for our simulation study where $K_{\text{max}}=6$ often becomes a binding upper optimization threshold. The choice of $K_{\text{max}}$ is crucial since it constrains the magnitude of possible omitted-variable biases in $\widehat{\beta}_K(t)$. The same issue applies to $\rho_{\text{min}}$ when optimizing the GCV in \eqref{gcv} over $\rho\in[\rho_{\text{min}},\rho_{\text{max}}]$ with $\rho_{\text{min}}\approx 0$. Therefore, we choose very conservative optimization intervals $[K_{\text{min}},K_{\text{max}}]=[1,150]$ and $[\rho_{\text{min}},\rho_{\text{max}}]=[10^{-6},200]$.

\subsection{Data Generating Processes and Simulation Results}\label{ssec:DGPandResults}
We consider five different Data Generating Processes (DGPs), as described in Table \ref{tab:dgp}. The DGPs Easy and Complicated represent a simple and a more complex version of Model \eqref{kneip}. The Complicated DGP is challenging due to the closeness of the PoI locations $\tau_1$ and $\tau_2$, which may trigger omitted-variable biases in $\widehat{\beta}(t)$ when omitting either $\tau_1$ or $\tau_2$. The two further DGPs NoPoI ($\calT=\emptyset$) and OnlyPoI ($\beta(t)\equiv 0$) are used to check the robustness of our PES-ES algorithm against model-misspecifications.
\begin{table}[!htbp]
\centering
\renewcommand{\arraystretch}{0.9}
\caption[DGP Definitions]{Data Generating Processes.} 
\begin{tabular}{lclccc}
\toprule
DGP  	      && \multicolumn{1}{c}{$\beta(t)$}  & $S$ & $\calT=\lbrace\tau_{1},\dots,\tau_{S}\rbrace$ & $\lbrace\beta_{1},\dots,\beta_{S}\rbrace$\\ 
\midrule
Easy  	    && $\beta(t) = -(t-1)^2+2 $ 	  & 2& $\lbrace 0.3, 0.6\rbrace$      & $\lbrace -3, 3   \rbrace $\\ 
Complicated && $\beta(t) = -5(t-0.5)^3-t+1$ & 3& $\lbrace 0.3, 0.4, 0.6\rbrace$ & $\lbrace -3, 3, 3\rbrace $\\
OnlyPoI     && $\beta(t)  \equiv 0$         & 2& $\lbrace 0.3, 0.6\rbrace$      & $\lbrace -3, 3\rbrace $\\
NoPoI  	    && $\beta(t) = -(t-1)^2+2 $ 	  & 0& $\emptyset$ 	                  & $\emptyset$\\ 
\bottomrule
\end{tabular}\label{tab:dgp}
\end{table}

For each DGP and two sample sizes ($n=250$ and $n=500$), we generate $1000$ replications of $n$ functions $X_i(t)$ observed at $p=300$ equidistant points $t_1, \ldots, t_p$ in $[0,1]$. In Appendix \ref{Appendix} we additionally present simulation results for $p=500$. The functions $X_i(t)$ are standard Brownian Motions, and the dependent variables $Y_i$ are generated according to Model \eqref{kneip} with $\epsilon_i \sim N(0, 0.125^2)$. Our simulation is implemented using the statistical language \textsf{R} \citep{rcore}, and the \textsf{R}-codes for reproducing the simulation results are part of the online supplement supporting this article \citep{LieRamRus19}.

\begin{table}[!ht]
\spacingset{1.1}
\centering
\tabcolsep=0.125cm 
\renewcommand{\arraystretch}{1.1}
\caption{Squared bias and variance of the estimators. Shades of gray show the ranking of the Mean Squared Error (MSE): lowest/highest MSE has the darkest/lightest gray-scale.}
\vspace{0.08cm}
\label{tab:sim}
\resizebox{\textwidth}{!}{%
  \newcommand{\myCell}[3]{\cellcolor{black!#1}\phantom{#2}#3}
\begin{tabular}{llcccccccccccc}
  \toprule&&
  \multicolumn{2}{c}{Easy} &&
  \multicolumn{2}{c}{Complicated} &&
  \multicolumn{2}{c}{NoPoI} &&
  \multicolumn{2}{c}{OnlyPoI}&\\
  \cmidrule{3-4}
  \cmidrule{6-7}
  \cmidrule{9-10}
  \cmidrule{12-13}
  \multicolumn{2}{l}{ $\int\widehat{\beta}(t)$}&
  \multicolumn{1}{c}{Bias$^2$} & \multicolumn{1}{c}{Var.} &&
  \multicolumn{1}{c}{Bias$^2$} & \multicolumn{1}{c}{Var.} &&
  \multicolumn{1}{c}{Bias$^2$} & \multicolumn{1}{c}{Var.}&&
  \multicolumn{1}{c}{Bias$^2$} & \multicolumn{1}{c}{Var.}&\\
  \multirow{5}{*}{\rotatebox{90}{$n=250$}}
  & PES
  & \myCell{40}{00}{0.02} & \myCell{40}{00}{0.22}
  && \myCell{20}{00}{0.21} & \myCell{20}{00}{1.98}
  && \myCell{15}{00}{0.00} & \myCell{15}{00}{0.02}
  && \myCell{20}{00}{0.00} & \myCell{20}{00}{0.08}\\
  & PES-ES
  & \myCell{30}{00}{0.02} & \myCell{30}{00}{0.24}
  && \myCell{30}{00}{0.16} & \myCell{30}{00}{1.68}
  && \myCell{30}{00}{0.00} & \myCell{30}{00}{0.01}
  && \myCell{40}{00}{0.00} & \myCell{40}{00}{0.06}\\
  & PES-2ES
  & \myCell{20}{00}{0.02} & \myCell{20}{00}{0.25}
  && \myCell{40}{00}{0.16} & \myCell{40}{00}{1.66}
  && \myCell{30}{00}{0.00} & \myCell{30}{00}{0.01}
  && \myCell{40}{00}{0.00} & \myCell{40}{00}{0.06}\\
  & KPS
  & \myCell{10}{00}{2.81} & \myCell{10}{0}{51.17}
  && \myCell{10}{}{155.17} & \myCell{10}{}{303.03}
  && \myCell{10}{00}{0.01} & \myCell{10}{00}{0.02}
  && \myCell{15}{00}{0.05} & \myCell{15}{00}{6.65}\\
  & CKS
  &  - & - 
  &&  - & - 
  && \myCell{30}{00}{0.00} & \myCell{30}{00}{0.00}
  &&  - & - \\
  \\[-0.9em]
  \multirow{5}{*}{\rotatebox{90}{$n=500$}}
  & PES
  & \myCell{20}{00}{0.01} & \myCell{20}{00}{0.06}
  && \myCell{20}{00}{0.05} & \myCell{20}{00}{0.55}
  && \myCell{20}{00}{0.00} & \myCell{20}{00}{0.01}
  && \myCell{30}{00}{0.00} & \myCell{30}{00}{0.01}\\
  & PES-ES
  & \myCell{40}{00}{0.00} & \myCell{40}{00}{0.05}
  && \myCell{40}{00}{0.04} & \myCell{40}{00}{0.38}
  && \myCell{20}{00}{0.00} & \myCell{20}{00}{0.01}
  && \myCell{30}{00}{0.00} & \myCell{30}{00}{0.01}\\
  & PES-2ES
  & \myCell{40}{00}{0.00} & \myCell{40}{00}{0.05}
  && \myCell{40}{00}{0.04} & \myCell{40}{00}{0.38}
  && \myCell{20}{00}{0.00} & \myCell{20}{00}{0.01}
  && \myCell{30}{00}{0.00} & \myCell{30}{00}{0.01}\\
  & KPS
  & \myCell{15}{00}{0.35} & \myCell{15}{0}{16.69}
  && \myCell{10}{0}{91.32} & \myCell{10}{}{245.88}
  && \myCell{10}{00}{0.01} & \myCell{10}{00}{0.01}
  && \myCell{15}{00}{0.00} & \myCell{15}{000}{0.5}\\
  & CKS
  &  - & - 
  &&  - & - 
  && \myCell{40}{00}{0.00} & \myCell{40}{00}{0.00}
  &&  - & - \\
  \\[-0.9em]
  \\[-0.7em]
  \multicolumn{2}{l}{ $\frac{1}{S}\sum\widehat{\beta}_s$}\\
  \multirow{4}{*}{\rotatebox{90}{$n=250$}}
  & PES
  & \myCell{20}{00}{0.01} & \myCell{20}{000}{0.1}
  && \myCell{20}{00}{0.01} & \myCell{20}{00}{0.09}
  &&  - & - 
  && \myCell{30}{00}{0.00} & \myCell{30}{00}{0.02}\\
  & PES-ES
  & \myCell{40}{00}{0.00} & \myCell{40}{00}{0.08}
  && \myCell{40}{00}{0.01} & \myCell{40}{00}{0.06}
  &&  - & - 
  && \myCell{30}{00}{0.00} & \myCell{30}{00}{0.02}\\
  & PES-2ES
  & \myCell{40}{00}{0.00} & \myCell{40}{00}{0.08}
  && \myCell{40}{00}{0.01} & \myCell{40}{00}{0.06}
  &&  - & - 
  && \myCell{30}{00}{0.00} & \myCell{30}{00}{0.02}\\
  & KPS
  & \myCell{15}{00}{0.03} & \myCell{15}{00}{0.54}
  && \myCell{15}{00}{1.59} & \myCell{15}{00000}{4}
  &&  - & - 
  && \myCell{15}{00}{0.00} & \myCell{15}{00}{0.06}\\
  \\[-0.9em]
  \multirow{4}{*}{\rotatebox{90}{$n=500$}}
  & PES
  & \myCell{20}{00}{0.00} & \myCell{20}{00}{0.02}
  && \myCell{30}{00}{0.00} & \myCell{30}{00}{0.01}
  &&  - & - 
  && \myCell{15}{00}{0.00} & \myCell{15}{00}{0.02}\\
  & PES-ES
  & \myCell{40}{00}{0.00} & \myCell{40}{00}{0.01}
  && \myCell{30}{00}{0.00} & \myCell{30}{00}{0.01}
  &&  - & - 
  && \myCell{40}{00}{0.00} & \myCell{40}{00}{0.00}\\
  & PES-2ES
  & \myCell{40}{00}{0.00} & \myCell{40}{00}{0.01}
  && \myCell{30}{00}{0.00} & \myCell{30}{00}{0.01}
  &&  - & - 
  && \myCell{40}{00}{0.00} & \myCell{40}{00}{0.00}\\
  & KPS
  & \myCell{15}{00}{0.01} & \myCell{15}{000}{0.2}
  && \myCell{15}{00}{0.78} & \myCell{15}{00}{2.92}
  &&  - & - 
  && \myCell{20}{00}{0.00} & \myCell{20}{00}{0.01}\\
  \\[-0.9em]
  \bottomrule
\end{tabular}
}
\end{table}
The upper panel of Table \ref{tab:sim} reports the integrated squared bias and the integrated variance for the estimator $\widehat{\beta}(t)$ of $\beta(t)$. The integrated squared bias is computed as $\int_0^1 (\bar{\beta}(t) - \beta(t))^2dt$, where $\bar{\beta}(t)=1000^{-1}\sum_{r=1}^{1000}\widehat{\beta}_r(t)$ is the mean of the estimates over all replications. The integrated variance is computed as $1000^{-1}\int_0^1 \sum_{r=1}^{1000}(\widehat{\beta}_r(t) - \bar{\beta}(t))^2dt$. The lower panel of Table \ref{tab:sim} reports the average squared bias $S^{-1}\sum_{s=1}^S(\bar{\beta}_s - \beta_s)^2$, with $\bar{\beta}_s=1000^{-1}\sum_{r=1}^{1000}\widehat{\beta}_{r,s}$, and the average variance $S^{-1}\sum_{s=1}^S1000^{-1}\sum_{r=1}^{1000}(\widehat{\beta}_{r,s} - \bar{\beta}_s)^2$ for the PoI coefficient estimators $\widehat{\beta}_s$, conditionally on the event that $\tau_s$ was correctly found\footnote{Note that it is impossible to compute estimation errors for non-found $\tau_s$.}, where a single $\tau_{s}$ is considered to be found correctly if $|\widehat{\tau}_s - \tau_{s}| < 0.01$. The latter requirement corresponds to an estimation precision of only $\pm 3$ grid points, which is substantially more challenging than the matching requirement originally used in \citet{KnePosSa2016}. The shades of gray in Table \ref{tab:sim} show the ranking of the mean squared error (MSE); the lowest/highest MSE (=Bias$^2$ + Var.) has the darkest/lightest gray-scale.

The simulation results for the slope parameters $\beta(t)$ and $\beta_1,\dots,\beta_S$ in the upper and lower panel of Table \ref{tab:sim} show that the smoothing-spline-based estimation algorithms PES and PES-ES clearly outperform the FPCA-based KPS estimator. The final ES-step in the PES-ES algorithm aims to remove further falsely selected point of impact candidates. This is advantageous in all DGPs, except for the Easy DGP with $n=250$ and the NoPoI DGP, where PES-ES and PES achieve essentially equivalent results. Note that the final ES-step is particularly advantageous for the Complicated DGP and the smaller sample size $n=250$, where KPS shows a very poor performance. Only in this particular case, one additional second repetition of the ES-step (PES-2ES) further reduces the variance. This improvement, however, is not substantial and does not justify the additionally involved computational burden of PES-2ES. PES-ES also performs very well in the NoPoI and the OnlyPoI DGPs, where PES-ES is actually a misspecified estimation procedure. In the case of NoPoI, PES-ES performs almost as well as the corresponding (minimax-optimal) benchmark-estimator CKS, and in the case of OnlyPoI, PES-ES is the best performing method.

\begin{table}[t]
\spacingset{1}
\centering
\tabcolsep=0.11cm
\caption{Percentage of replications with correct detection of all points of impact $\tau_1, \ldots, \tau_S$.}
\vspace{0.08cm}
\label{tab:sim2}
\tabcolsep=0.11cm
\newcommand{\myCell}[3]{\cellcolor{black!#1}\phantom{#2}#3}
 \begin{tabular}{llccccccccccccc}
\toprule
&&\multicolumn{5}{c}{300 grid points} &&& \multicolumn{5}{c}{500 grid points}\\
\cmidrule{3-7}\cmidrule{10-14}
&&\multicolumn{1}{c}{Easy} &&
\multicolumn{1}{c}{Compl.}  &&
\multicolumn{1}{c}{OnlyPoI}
&&&\multicolumn{1}{c}{Easy} &&
\multicolumn{1}{c}{Compl.}  &&
\multicolumn{1}{c}{OnlyPoI}&\\
\midrule
\multirow{4}{*}{\rotatebox{90}{$n=250$}}
& PES
& \myCell{15}{}{97.5}
&& \myCell{15}{}{77.4}
&& \myCell{15}{}{99}&
&& \myCell{15}{}{97}
&& \myCell{15}{}{83.8}
&& \myCell{15}{}{99.1}\\
& PES-ES
& \myCell{40}{}{97.6}
&& \myCell{40}{}{79.3}
&& \myCell{40}{}{99.2}&
&& \myCell{40}{}{97.3}
&& \myCell{20}{}{85.3}
&& \myCell{40}{}{99.2}\\
& PES-2ES
& \myCell{40}{}{97.6}
&& \myCell{40}{}{79.3}
&& \myCell{40}{}{99.2}&
&& \myCell{40}{}{97.3}
&& \myCell{40}{}{85.8}
&& \myCell{40}{}{99.2}\\
& KPS
& \myCell{10}{}{89.7}
&& \myCell{10}{}{19.3}
&& \myCell{10}{}{98.7}&
&& \myCell{10}{}{89.5}
&& \myCell{10}{}{24}
&& \myCell{10}{}{98.5}\\
\\[-0.9em]
\multirow{4}{*}{\rotatebox{90}{$n=500$}}
& PES
& \myCell{15}{}{99.3}
&& \myCell{15}{}{94.6}
&& \myCell{15}{}{99.9}&
&& \myCell{40}{}{99.3}
&& \myCell{15}{}{94}
&& \myCell{20}{}{99.9}\\
& PES-ES
& \myCell{40}{}{99.4}
&& \myCell{20}{}{95.7}
&& \myCell{15}{}{99.9}&
&& \myCell{20}{}{99.2}
&& \myCell{40}{}{95.3}
&& \myCell{20}{}{99.9}\\
& PES-2ES
& \myCell{40}{}{99.4}
&& \myCell{40}{}{95.8}
&& \myCell{15}{}{99.9}&
&& \myCell{20}{}{99.2}
&& \myCell{40}{}{95.3}
&& \myCell{20}{}{99.9}\\
& KPS
& \myCell{10}{}{96.9}
&& \myCell{10}{}{37.2}
&& \myCell{40}{}{100}&
&& \myCell{10}{}{97}
&& \myCell{10}{}{41.9}
&& \myCell{10}{}{99.5}\\
\\[-0.9em]
\bottomrule
\end{tabular}

\end{table}

Table \ref{tab:sim2} reports for each estimator and sample size the percentage of replications where all PoI locations $\tau_1,\dots,\tau_S$ are found correctly. The left part of the table contains the results for functions observed on $p=300$ grid points and the right part for $p=500$ grid points. PES-ES and PES-2ES outperform all competitors, except in the case of OnlyPoI with $n=500$, where all estimation procedures show essentially the same performance. Again, the difference between PES(-(2)ES) and KPS is particularly large for the smaller sample size $n=250$ and the Complicated DGP. Increasing the resolution of the grid from $p=300$ to $p=500$ does not change the results. Similarly, the increased resolution also does not affect the precision of the estimate for the slope parameter $\beta(\cdot)$ and $\beta_1,\dots,\beta_s$, see Table \ref{tab:sim_p500} in Appendix \ref{Appendix}.

To show the performance boost of using standardized data for locating the potential PoIs (as described at the end of Section \ref{ssec:presel}), we report the simulation results without standardizing the data (see Tables \ref{tab:sim_noscale_p300} and \ref{tab:sim_noscale_p500} in Appendix \ref{Appendix}). The results show that the standardization of the data is beneficial for the Complicated DGP. Table \ref{tab:sim_noise} in Appendix \ref{Appendix} shows the simulation results for the Complicated DGP, but with different noise-to-signal ratios, that is, with different values for the error variance in model \eqref{kneip}. PES(-ES) still outperforms KPS significantly; however, it turns out that the difference becomes less pronounced as the noise-to-signal ratio increases.

\section{Application}\label{sec:app}
To illustrate the practical importance of the functional linear regression model with points of impact, we present an application to data from Google AdWords, which is the most popular online advertising platform and of fundamental importance for Alphabet's (Google's parent company) economic success (in 2014, $90$ percent of Alphabet's sales came from AdWords). Online advertising, in turn, is the most important branch of today's advertising industry, with an expected U.S.~revenue of $60$ billion USD in 2016 \citep{PWCADRevReport2016_1}. The case study described below is motivated by the needs of \textit{Crealytics}, the company that generously provided the data. Today this company uses the described method---with some further confidential enhancements---to support their daily business.

The main pricing mechanism at Google AdWords is the so-called Pay-Per-Click (PPC) mechanism. Here, advertisers (e.g., an online outdoor shop in our application) can bid for a sponsored ``impression'' to be displayed along with Google's search results when a user conducts a search query related to a specific keyword (e.g., \texttt{outdoor jacket})\footnote{Sponsored impressions link to the advertised homepage---they are similar to, but distinguishable from ordinary Google search results.}. The basic building block of an online ad campaign is a text corpus of (hundreds, thousands, or ten-thousands of, etc.) keywords related to the advertised products.

The limited number of sponsored impressions is allocated by an auction. Advertisers whose impression appears on the display are chosen according to their ad-rank, which is basically their original bid, i.e., the maximum ``costs-per-click'' an advertiser is willing to pay times the quality score, a discrete metric (from $1$, the lowest, to $10$, the best) determining the relevance of an advertiser's impression. Google AdWords auctions are time continuous and an advertiser only pays if a user clicks on the displayed impression. (See \citeauthor{Ged2014}, \citeyear{Ged2014}, for an in-depth introduction to Google AdWords.) 

The bidding process is usually based on bidding softwares that evaluate specific key-figures. One of the most important key-figures is the so-called Click-Though Rate (CTR), which is defined as the daily number of clicks per impression. The CTR estimates the current probability of receiving a click on a sponsored impression and therefore plays an important role in assisting the bidding process on a short-term basis \citep{Ged2014}.

The economic success of ad campaigns, however, also depends on long-sighted bidding strategies taking into account product specific (time-global) seasonalities as well as (time-local) events, such as the importance of Valentine's Day for an online flower shop. Unfortunately, existing key-figures such as the CTR only provide a daily perspective and are not suitable for assisting in the implementation of long-sighted bidding strategies.
Therefore, the functional linear regression model with points of impact is a suitable methodology to identify the (global and local) functional relationship between the \emph{yearly} clicks and the \emph{yearly} trajectories of daily impressions---leading to a long-sighted version of the CTR.

\begin{figure}
\centering
\includegraphics[width=\textwidth]{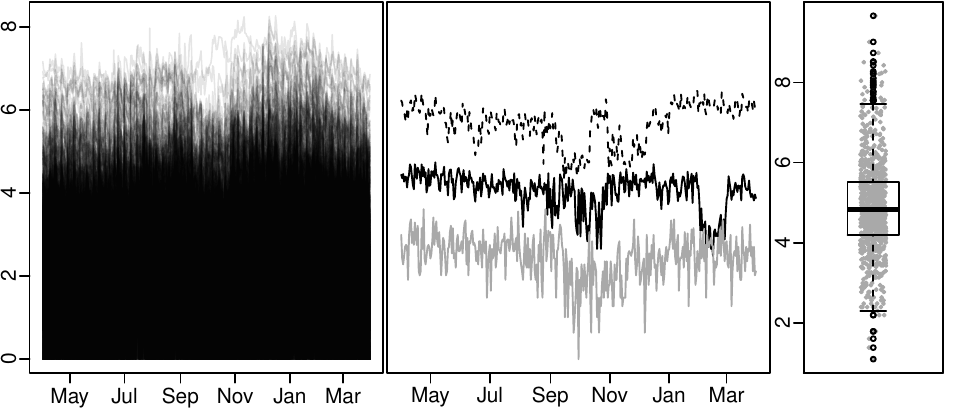}
\caption[Log Daily Impressions and Exemplary Salewa Curves]{\textsc{Left:} Yearly trajectories of daily logarithmized numbers of impressions. \textsc{Middle:} Three exemplary trajectories $X_i(t)$. \textsc{Right:} Logarithmized yearly clicks $Y_i$.}\label{fig:impressions}
\end{figure}

As a yearly measure of clicks, we use the logarithmized yearly sums of clicks, i.e., $Y_i=\log(C_i)$ with $C_i:=\sum_{t=1}^{365}\texttt{clicks}_{it}$, where $i$ indexes the $i$th keyword of the considered ad campaign. As a yearly measure of impressions, we use the yearly trajectories of daily logarithmized numbers of impressions, i.e., $X_i(t)=\log(\mathcal{I}_i(t))$ with $\mathcal{I}_i(t):=\texttt{impressions}_{it}$, where $t=1,\dots,365$ indexes the days of the considered year. Our application uses data from a real Google AdWords campaign run by an online store selling outdoor equipment in the year from April 1st, 2012 to March 31st, 2013. The left plot in Figure \ref{fig:impressions} shows all trajectories $X_i(t)$ of the considered ad campaign. The middle plot shows three exemplary (logarithmized) impression trajectories $X_i(t)$.
The right panel shows the (logarithmized) yearly sum of clicks $Y_i$, received on the impressions of the $i$th keyword.

The data are provided by \textit{Crealytics} (\href{https://crealytics.com/what-we-offer/}{www.crealytics.com}), an online advertising service provider with offices in Berlin (Germany), London (UK), and New York City (USA). The considered ad campaign is that of an online store selling outdoor equipment. (For reasons of confidentiality, we cannot publish the company's name). A lot of keywords received no impression during the considered time span of 365 days from April 1st, 2012, to March 31st, 2013. Therefore, we consider only the well established and relevant keywords that have been used on at least 320 days within the considered time span---leading to $n=903$ trajectories observed at $p=365$ grid points. The very few missing values in the logarithmized impression trajectories are imputed by zeros since a missing value means that the corresponding keyword did not receive an impression.

The considered functional linear regression model with PoIs in \eqref{kneip} is identifiable if the covariance function of the function-valued explanatory variable $X_i$ is sufficiently non-smooth at the diagonal \citep[see Section \ref{ssec:KPS} and Theorem 3 in][]{KnePosSa2016}. \cite{KnePosSa2016} propose the following consistent estimator $\widehat\kappa$ for their $\kappa$ controlling the smoothness at the diagonal of the covariance function:
\begin{equation*}
\widehat\kappa = \log_2\left(\frac{(1/(p-2k_\delta))\sum_{j \in \calJ_{0,\delta}}\sum_{i=1}^n Z_{\delta,X_i}(t_j)^2}{(1/(p-2k_\delta)) \sum_{j \in \calJ_{0,\delta}}\sum_{i=1}^n Z_{\delta/2,X_i}(t_j)^2} \right).
\end{equation*}
An estimate of $\widehat\kappa< 2$ indicates identifiability, which is clearly fulfilled in our case where $\widehat{\kappa} = 0.03$.

\begin{figure}[t]
\centering
\hspace{2.2cm}
\includegraphics[width=\textwidth]{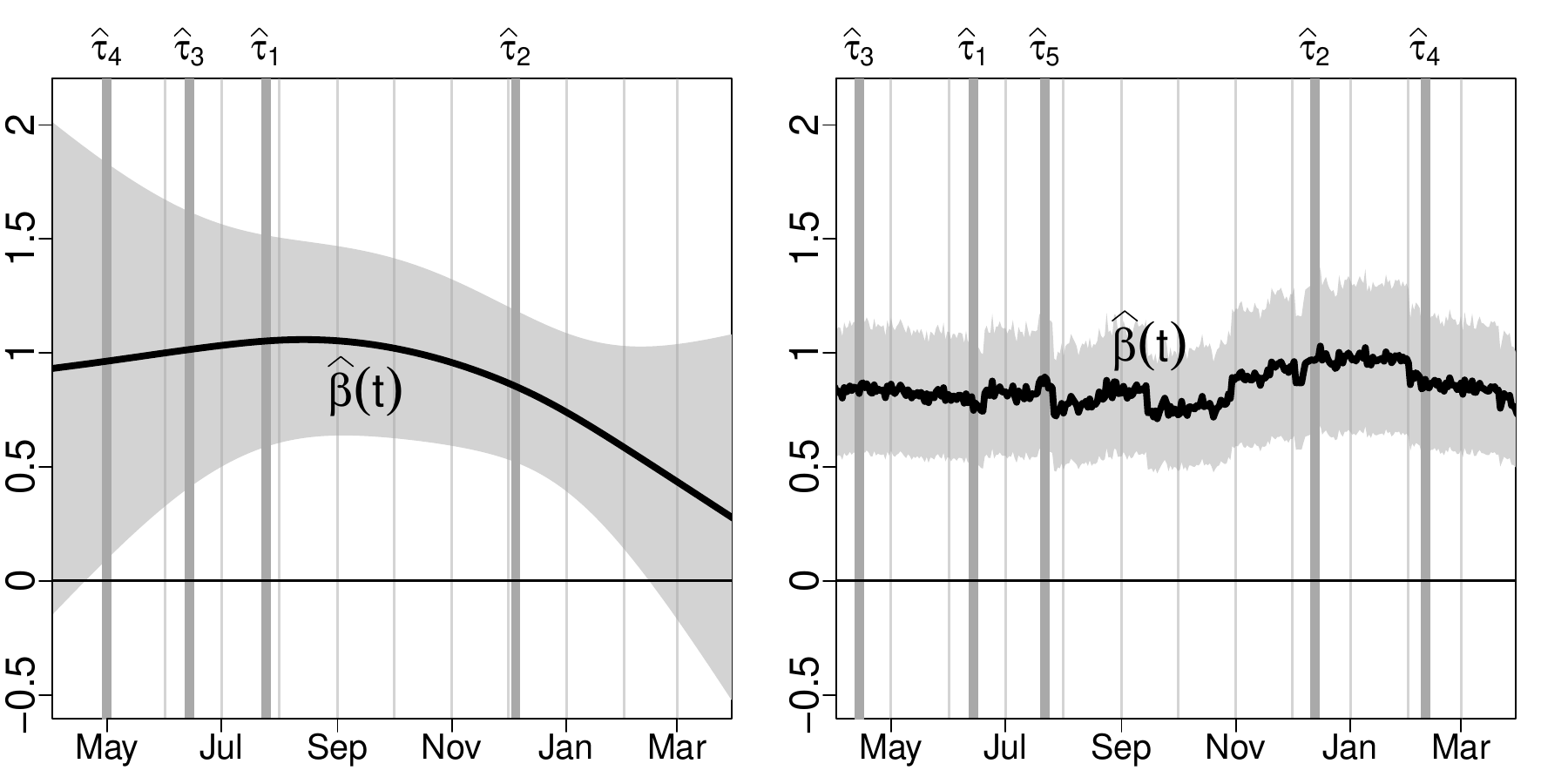}
\caption[AdWords PES-ES Estimation - Summary]{Result of the PES-ES (left panel) and KPS (right panel) estimate for $\beta(\cdot)$. The variabilities of the estimators are visualized using the gray shaded bands (see Remark 1).}
\label{fig:estimates}
\end{figure}

\begin{table}[t]
  \spacingset{1}
  \centering
  \tabcolsep=0.11cm
  \caption{Estimate of PoI parameters $\beta_r$}
  \label{tab:PoI_coefs}
  \begin{tabular}{llccllc}
    \toprule
    \multicolumn{3}{c}{PES-ES} && \multicolumn{3}{c}{KPS}\\
    \cmidrule{1-3}\cmidrule{5-7}
    Location & Coef. & St.Err. & & Location & Coef. & St.Err.\\
    \midrule
    ($\widehat{\tau}_4$) May 01   &          -0.17$^{***}$ & 0.04 && ($\widehat{\tau}_3$) April 14 &           -0.10$^{**}$  & 0.03\\
    ($\widehat{\tau}_3$) June 14  &\phantom{-}0.22$^{***}$ & 0.03 && ($\widehat{\tau}_1$) June 14  & \phantom{-}0.22$^{***}$ & 0.03\\
    ($\widehat{\tau}_1$) July 25  &          -0.15$^{***}$ & 0.03 && ($\widehat{\tau}_5$) July 22  &           -0.17$^{***}$ & 0.03\\
    ($\widehat{\tau}_2$) December 05 &  \phantom{-}0.01    & 0.03 && ($\widehat{\tau}_2$) December  13 & \phantom{-}0.06$^{*}$ & 0.03\\
                                  &                        &      && ($\widehat{\tau}_4$) February 10 & -0.11$^{***}$ & 0.03\\
    \bottomrule
  \end{tabular}
\end{table}

The estimation results from applying our PES-ES estimation algorithm and the originally proposed KPS procedure are summarized in Figure \ref{fig:estimates} and Table \ref{tab:PoI_coefs}. In case of the PES-ES estimate, the function-valued slope parameter $\widehat{\beta}(t)$ shows a peak in the late summer and a pronounced negative trend towards the end of the considered period. The shape of $\widehat{\beta}(t)$ is in line with our expectations since the demand for outdoor equipment is generally greater during the summer months than during the winter months. The negative trend towards the end of the considered period is due to the strongly increased competition for outdoor equipment ads in Google AdWords during the considered period. Additionally, the estimation procedure identifies four PoIs (in order of the magnitude of $|\widehat\beta_s|$): June 14th ($\widehat{\tau}_3$; $\widehat\beta_3=0.22$), May 1st ($\widehat{\tau}_4$; $\widehat\beta_4=-0.17$), July 25th ($\widehat{\tau}_1$; $\widehat\beta_1=-0.15$), and December 5th ($\widehat{\tau}_2$; $\widehat\beta_2=0.01$), where the effect of the PoI at $\widehat{\tau}_2$ seems to be of lower importance.

\begin{remark}
Drawing inference about the function-valued slope coefficient and the PoI parameters is a difficult issue in regression models with functional predictors. This is due to the fact that estimation in such models involves an ill-posed inversion problem and the estimator of the function-valued slope parameter is not asymptotically normal in the strong topology \citep{CarMasSar2007}. In addition, it is difficult to construct confidence regions for random elements in infinite dimensional Hilbert spaces with proper coverage probability \citep{ChoRei2018}. All we can do is to visualize the variability of the estimator that is due to the error term $\epsilon_i$. For this purpose, we approximate the sampling variance of the composite parameter vector $\beta_{\widetilde{\mathcal{T}}}^\rho$ using Eq.~(15.16) in \cite{RamSil2005}, Ch.~15, and show Bonferroni-adjusted Gaussian (invalid) confidence intervals in Figure \ref{fig:estimates}. 
\end{remark}

The PoI $\widehat{\tau}_3$ on June 14th, with coefficient $\widehat{\beta}_3=0.22$, summarizes two positive effects. On the one hand, the store started a contest on May 23rd, 2012, giving away outdoor gear. This contest ended on June 13th, i.e., one day before the PoI which  resulted in an increased click-through ratio of contest participants looking for the winners. On the other hand, the closest competitor started the spring sale, which led to a spillover bringing many interested buyers onto the homepage to compare prices. 

The two other significant PoIs are explained by effects specific to the German calendar (about 80 percent of the customers live in Germany). The PoI $\widehat{\tau}_4$ on May 1st, with coefficient $\widehat{\beta}_4=-0.17$, marks the Labor Day (commemorating the Haymarket Riot in Chicago in 1886), a national holiday in Germany which is typically an opportunity for family outings. Similar in interpretation, the PoI $\widehat{\tau}_1$ on July 25th, with coefficient $\widehat{\beta}_1=-0.15$, marks the beginning of the official summer holidays in Baden-W\"urttemberg and Lower Saxony---two large German states. Both PoIs show a negative sign, which is due to a higher volume in search queries related to outdoor activities, however, the users do not click on the sponsored impressions since they do not intend to buy something---they are only searching the Internet for (free) information on hiking trails etc., which results in a lower CTR.

By contrast to the PES-ES estimate of $\beta(\cdot)$, the KPS estimate of $\beta(\cdot)$ is difficult to interpret and does not fit to our expectations (see right panel of Fig.~\ref{fig:estimates}): the trajectory is very unstable, does not show the expected peak in late summer, and does not show the plausible  negative trend towards the end of the considered period. Regarding the PoI selections, the KPS approach identifies essentially the same PoIs as the PES-ES approach, but favors one additional PoI at February 10 (see Table \ref{tab:PoI_coefs}), which has a significant negative impact ($\widehat{\beta}_4=-0.11$) on the outcome variable. This additional PoI may reflect a compensation for the missing negative trend in the KPS estimate of $\beta(\cdot)$; see our discussion in Section \ref{ssec:estadj}.

The log-transformations in $Y_i=\log(C_i)$ and $X_i(t)=\log(\mathcal{I}_i(t))$ allow us to interpret the estimated slope coefficients as elasticities. Taking derivatives with respect to $\mathcal{I}_i(t)$ at a single time point $t$ leads to the following \emph{time-local} elasticity:
\begin{equation*}
\frac{\%\Delta C_i}{\%\Delta \mathcal{I}_i(t)}\approx\left\{\begin{array}{ll}\widehat\beta_s&\text{if}\quad t=\widehat{\tau}_s\\
0&\text{else}.\end{array}\right.
\end{equation*}
That is, time-local changes in $\mathcal{I}_i(t)$ generally have no (i.e., practically negligible) effects on the yearly clicks $C_i$, except at PoIs, i.e., if  $t=\widehat\tau_1,\dots,\widehat\tau_{\widehat S}$. For instance, a $1\%$ increase in the impressions at the time point of the after-contest PoI ($t=\widehat{\tau}_3$) causes (on average) a $0.22\%$ ($\widehat\beta_3= 0.22$) increase in the yearly clicks. 

The function-valued slope parameter $\widehat{\beta}(t)$ does not contribute to the time-local elasticities; however, it determines the elasticities with respect to time-global changes in the impressions, for instance, over the course of a month. The following Riemann sum allows for a simple, approximative approach to interpret such \emph{time-global} elasticities: 
\begin{equation*}
\widehat{\log(C_i)}\approx\frac{1}{365}\sum_{t=1}^{365}\widehat{\beta}(t)\log\big(\mathcal{I}_i(t)\big)+\sum_{s=1}^{\widehat{S}}\widehat{\beta}_s\log\big(\mathcal{I}_i(\widehat{\tau}_s)\big).
\end{equation*}
For instance, the total elasticity of $C_i$ with respect to $\mathcal{I}_i(t)$ for \emph{all} $t\in\text{August}$ is given by
\begin{equation*}
\sum_{t\in\text{August}}\frac{\%\Delta C_i}{\%\Delta \mathcal{I}_i(t)}\approx\frac{1}{365}\sum_{t\in\text{August}}\widehat{\beta}(t)+\sum_{s=1}^{\widehat{S}}\widehat{\beta}_s\mathbf{1}_{(\widehat{\tau}_s\in\text{August})},
\end{equation*}
where $\mathbf{1}_{(\texttt{TRUE})}=1$ and $\mathbf{1}_{(\texttt{FALSE})}=0$. That is, a 1\% increase in the impressions $\mathcal{I}_i(t)$, simultaneously for all $t\in\text{August}$, causes a 0.1\% increase in the yearly clicks since $365^{-1}\sum_{t\in\text{August}}\widehat{\beta}(t)+\sum_{s=1}^{\widehat{S}}\widehat\beta_s\mathbf{1}_{(\widehat{\tau}_s\in\text{August})}\approx 0.1$. Hence, the time-global August-elasticity is half the size of the elasticity of the after-contest PoI. This is absolutely plausible since the super-imposed influence of the contest and the spillover definitely outperforms a high-season month such as August in terms of clicks-per-impressions.

\section{Conclusion} \label{sec:con}
In this work we propose an improved algorithm for estimating the unknown model components of the functional linear regression model with points of \cite{KnePosSa2016}. Our estimation algorithm decouples the estimation of the points of impact from the estimation of the function-valued slope parameter. The first step of the estimation algorithm, allows for a consistent estimation of the points of impact without knowledge (or pre-estimation) of the slope function. Given the consistent estimates of the points of impact, the second step of the estimation algorithm consists of an essentially classical estimation of the function-valued slope parameter. For this latter step we propose a generalization of the penalized smoothing splines estimator of \cite{CraKneSa2009}, which allows to incorporate the estimates of the points of impacts. A further minor finite sample improvement is achieved by repeating the estimation of the points of impact, given the estimate of the function-valued slope parameter from the second step and by a finial repetition of the estimation of the slope parameter, given the updated estimates of the points of impact.

The new estimation algorithm significantly improves the original estimation procedure by \cite{KnePosSa2016}. Using an extensive simulation study, we assess the robustness of our estimation algorithm for different data generating processes, different signal-to-noise ratios, different sample sizes and different sampling resolutions for discretizing the function-valued predictors.

The paper was originally motivated by an interesting case study on a Google AdWords ad campaign. Our proposed functional linear regression model with points of impacts allows for data-based insights into the (time-global) seasonal factors and the (time-local) events influencing the yearly number of clicks on impressions of the considered Google AdWords online ad campaign.

\newpage
\appendix

\section{Additional simulation setups}\label{Appendix}

\begin{table}[!ht]
\spacingset{1.1}
\centering
\tabcolsep=0.125cm 
\renewcommand{\arraystretch}{1.1}
\caption{Squared bias and variance of the estimators. Lowest/highest MSE has the darkest/lightest gray-scale. \textbf{Scenario:} No standardization of the functions in preselection step and $p=300$ grid points.}
\vspace{0.08cm}
\label{tab:sim_noscale_p300}
\resizebox{\textwidth}{!}{%
  \newcommand{\myCell}[3]{\cellcolor{black!#1}\phantom{#2}#3}
\begin{tabular}{llccccccccccccccc}
\toprule&&
  \multicolumn{2}{c}{Easy} &&
  \multicolumn{2}{c}{Complicated} &&
  \multicolumn{2}{c}{NoPoI} &&
  \multicolumn{2}{c}{OnlyPoI}&\\
  \cmidrule{3-4}
  \cmidrule{6-7}
  \cmidrule{9-10}
  \cmidrule{12-13}
  \multicolumn{2}{l}{ $\int\widehat{\beta}(t)$}&
  \multicolumn{1}{c}{Bias$^2$} & \multicolumn{1}{c}{Var.} &&
  \multicolumn{1}{c}{Bias$^2$} & \multicolumn{1}{c}{Var.} &&
  \multicolumn{1}{c}{Bias$^2$} & \multicolumn{1}{c}{Var.}&&
  \multicolumn{1}{c}{Bias$^2$} & \multicolumn{1}{c}{Var.}&\\
  \multirow{5}{*}{\rotatebox{90}{$n=250$}}
& PES
& \myCell{15}{00}{0.05} & \myCell{15}{00}{0.89}
&& \myCell{15}{00}{2.02} & \myCell{15}{0}{12.36}
&& \myCell{15}{00}{0.00} & \myCell{15}{00}{0.02}
&& \myCell{15}{00}{0.00} & \myCell{15}{00}{0.08}\\
& PES-ES
& \myCell{20}{00}{0.05} & \myCell{20}{00}{0.74}
&& \myCell{40}{00}{1.81} & \myCell{40}{0}{11.64}
&& \myCell{40}{00}{0.00} & \myCell{40}{00}{0.01}
&& \myCell{40}{00}{0.00} & \myCell{40}{00}{0.07}\\
& PES-2ES
& \myCell{40}{00}{0.04} & \myCell{40}{00}{0.72}
&& \myCell{20}{00}{1.85} & \myCell{20}{0}{11.69}
&& \myCell{40}{00}{0.00} & \myCell{40}{00}{0.01}
&& \myCell{40}{00}{0.00} & \myCell{40}{00}{0.06}\\
& KPS
& \myCell{10}{00}{3.98} & \myCell{10}{0}{60.37}
&& \myCell{10}{}{139.62} & \myCell{10}{}{301.13}
&& \myCell{10}{00}{0.01} & \myCell{10}{00}{0.02}
&& \myCell{10}{00}{0.12} & \myCell{10}{0}{10.97}\\
\\[-0.9em]
\multirow{5}{*}{\rotatebox{90}{$n=500$}}
& PES
& \myCell{15}{00}{0.01} & \myCell{15}{00}{0.23}
&& \myCell{15}{00}{0.87} & \myCell{15}{00}{6.04}
&& \myCell{20}{00}{0.00} & \myCell{20}{00}{0.01}
&& \myCell{40}{00}{0.00} & \myCell{40}{00}{0.01}\\
& PES-ES
& \myCell{20}{00}{0.01} & \myCell{20}{000}{0.2}
&& \myCell{20}{00}{0.89} & \myCell{20}{00}{5.55}
&& \myCell{20}{00}{0.00} & \myCell{20}{00}{0.01}
&& \myCell{20}{00}{0.00} & \myCell{20}{00}{0.02}\\
& PES-2ES
& \myCell{40}{00}{0.01} & \myCell{40}{00}{0.19}
&& \myCell{40}{000}{0.9} & \myCell{40}{00}{5.46}
&& \myCell{20}{00}{0.00} & \myCell{20}{00}{0.01}
&& \myCell{20}{00}{0.00} & \myCell{20}{00}{0.02}\\
& KPS
& \myCell{10}{00}{0.69} & \myCell{10}{0}{23.39}
&& \myCell{10}{0}{82.42} & \myCell{10}{}{241.99}
&& \myCell{10}{00}{0.01} & \myCell{10}{00}{0.01}
&& \myCell{10}{00}{0.01} & \myCell{10}{00}{1.77}\\
\\[-0.9em]
\\[-0.7em]
\multicolumn{2}{l}{ $\frac{1}{S}\sum\widehat{\beta}_s$}\\
\multirow{4}{*}{\rotatebox{90}{$n=250$}}
& PES
& \myCell{15}{00}{0.02} & \myCell{15}{00}{0.55}
&& \myCell{15}{00}{0.06} & \myCell{15}{00}{0.42}
&&  - & - 
&& \myCell{15}{00}{0.00} & \myCell{15}{00}{0.02}\\
& PES-ES
& \myCell{20}{00}{0.02} & \myCell{20}{00}{0.46}
&& \myCell{20}{00}{0.04} & \myCell{20}{00}{0.33}
&&  - & - 
&& \myCell{40}{00}{0.00} & \myCell{40}{00}{0.02}\\
& PES-2ES
& \myCell{40}{00}{0.02} & \myCell{40}{00}{0.45}
&& \myCell{40}{00}{0.04} & \myCell{40}{00}{0.32}
&&  - & - 
&& \myCell{40}{00}{0.00} & \myCell{40}{00}{0.02}\\
& KPS
& \myCell{10}{00}{0.04} & \myCell{10}{00}{0.65}
&& \myCell{10}{00}{1.01} & \myCell{10}{00}{3.31}
&&  - & - 
&& \myCell{10}{00}{0.00} & \myCell{10}{00}{0.14}\\
\\[-0.9em]
\multirow{4}{*}{\rotatebox{90}{$n=500$}}
& PES
& \myCell{15}{00}{0.00} & \myCell{15}{000}{0.1}
&& \myCell{15}{00}{0.03} & \myCell{15}{00}{0.14}
&&  - & - 
&& \myCell{20}{00}{0.00} & \myCell{20}{00}{0.00}\\
& PES-ES
& \myCell{40}{00}{0.00} & \myCell{40}{00}{0.09}
&& \myCell{20}{00}{0.02} & \myCell{20}{00}{0.11}
&&  - & - 
&& \myCell{20}{00}{0.00} & \myCell{20}{00}{0.00}\\
& PES-2ES
& \myCell{40}{00}{0.00} & \myCell{40}{00}{0.09}
&& \myCell{40}{00}{0.02} & \myCell{40}{000}{0.1}
&&  - & - 
&& \myCell{20}{00}{0.00} & \myCell{20}{00}{0.00}\\
& KPS
& \myCell{10}{00}{0.01} & \myCell{10}{00}{0.22}
&& \myCell{10}{00}{0.49} & \myCell{10}{00}{2.25}
&&  - & - 
&& \myCell{10}{00}{0.00} & \myCell{10}{00}{0.02}\\
\\[-0.9em]
\bottomrule
\end{tabular}
}
\end{table}

\newpage

\begin{table}[!ht]
\spacingset{1.1}
\centering
\tabcolsep=0.125cm 
\renewcommand{\arraystretch}{1.1}
\caption{Squared bias and variance of the estimators. Lowest/highest MSE has the darkest/lightest gray-scale. \textbf{Scenario:} With standardization of the functions in preselection step and $p=500$ grid points.}
\vspace{0.08cm}
\label{tab:sim_p500}
\resizebox{\textwidth}{!}{%
  \newcommand{\myCell}[3]{\cellcolor{black!#1}\phantom{#2}#3}
\begin{tabular}{llcccccccccccc}
\toprule&&
  \multicolumn{2}{c}{Easy} &&
  \multicolumn{2}{c}{Complicated} &&
  \multicolumn{2}{c}{NoPoI} &&
  \multicolumn{2}{c}{OnlyPoI}&\\
  \cmidrule{3-4}
  \cmidrule{6-7}
  \cmidrule{9-10}
  \cmidrule{12-13}
  \multicolumn{2}{l}{ $\int\widehat{\beta}(t)$}&
  \multicolumn{1}{c}{Bias$^2$} & \multicolumn{1}{c}{Var.} &&
  \multicolumn{1}{c}{Bias$^2$} & \multicolumn{1}{c}{Var.} &&
  \multicolumn{1}{c}{Bias$^2$} & \multicolumn{1}{c}{Var.}&&
  \multicolumn{1}{c}{Bias$^2$} & \multicolumn{1}{c}{Var.}&\\
  \multirow{5}{*}{\rotatebox{90}{$n=250$}}
  & PES
  & \myCell{20}{00}{0.04} & \myCell{20}{00}{0.37}
  && \myCell{20}{00}{0.16} & \myCell{20}{00}{1.43}
  && \myCell{10}{00}{0.01} & \myCell{10}{00}{0.04}
  && \myCell{20}{00}{0.00} & \myCell{20}{00}{0.06}\\
  & PES-ES
  & \myCell{40}{00}{0.03} & \myCell{40}{000}{0.3}
  && \myCell{30}{00}{0.09} & \myCell{30}{00}{0.94}
  && \myCell{30}{00}{0.00} & \myCell{30}{00}{0.02}
  && \myCell{30}{00}{0.00} & \myCell{30}{00}{0.04}\\
  & PES-2ES
  & \myCell{30}{00}{0.04} & \myCell{30}{000}{0.3}
  && \myCell{40}{00}{0.08} & \myCell{40}{00}{0.94}
  && \myCell{30}{00}{0.00} & \myCell{30}{00}{0.02}
  && \myCell{40}{00}{0.00} & \myCell{40}{00}{0.03}\\
  & KPS
  & \myCell{10}{00}{2.62} & \myCell{10}{0}{46.83}
  && \myCell{10}{}{135.19} & \myCell{10}{}{288.08}
  && \myCell{15}{00}{0.01} & \myCell{15}{00}{0.02}
  && \myCell{15}{00}{0.09} & \myCell{15}{00}{8.14}\\
  & CKS
  &  - & - 
  &&  - & - 
  && \myCell{40}{00}{0.00} & \myCell{40}{00}{0.01}
  &&  - & - \\
  \\[-0.9em]
  \multirow{5}{*}{\rotatebox{90}{$n=500$}}
  & PES
  & \myCell{20}{00}{0.01} & \myCell{20}{00}{0.09}
  && \myCell{20}{00}{0.06} & \myCell{20}{00}{0.41}
  && \myCell{15}{00}{0.00} & \myCell{15}{00}{0.02}
  && \myCell{40}{00}{0.00} & \myCell{40}{00}{0.01}\\
  & PES-ES
  & \myCell{40}{00}{0.01} & \myCell{40}{00}{0.08}
  && \myCell{30}{00}{0.05} & \myCell{30}{00}{0.38}
  && \myCell{30}{00}{0.00} & \myCell{30}{00}{0.01}
  && \myCell{30}{00}{0.00} & \myCell{30}{00}{0.02}\\
  & PES-2ES
  & \myCell{40}{00}{0.01} & \myCell{40}{00}{0.08}
  && \myCell{40}{00}{0.04} & \myCell{40}{00}{0.38}
  && \myCell{30}{00}{0.00} & \myCell{30}{00}{0.01}
  && \myCell{30}{00}{0.00} & \myCell{30}{00}{0.02}\\
  & KPS
  & \myCell{15}{00}{0.43} & \myCell{15}{0}{17.82}
  && \myCell{10}{0}{91.76} & \myCell{10}{0}{238.1}
  && \myCell{15}{00}{0.01} & \myCell{15}{00}{0.01}
  && \myCell{15}{00}{0.01} & \myCell{15}{00}{2.71}\\
  & CKS
  &  - & - 
  &&  - & - 
  && \myCell{40}{00}{0.00} & \myCell{40}{00}{0.00}
  &&  - & - \\
  \\[-0.9em]
  \\[-0.7em]
  \multicolumn{2}{l}{ $\frac{1}{S}\sum\widehat{\beta}_s$}\\
  \multirow{4}{*}{\rotatebox{90}{$n=250$}}
  & PES
  & \myCell{40}{00}{0.01} & \myCell{40}{00}{0.08}
  && \myCell{20}{00}{0.01} & \myCell{20}{00}{0.02}
  &&  - & - 
  && \myCell{40}{00}{0.00} & \myCell{40}{00}{0.04}\\
  & PES-ES
  & \myCell{20}{00}{0.01} & \myCell{20}{00}{0.11}
  && \myCell{40}{00}{0.01} & \myCell{40}{00}{0.02}
  &&  - & - 
  && \myCell{30}{00}{0.00} & \myCell{30}{00}{0.05}\\
  & PES-2ES
  & \myCell{30}{00}{0.01} & \myCell{30}{00}{0.11}
  && \myCell{40}{00}{0.01} & \myCell{40}{00}{0.02}
  &&  - & - 
  && \myCell{30}{00}{0.00} & \myCell{30}{00}{0.05}\\
  & KPS
  & \myCell{15}{00}{0.03} & \myCell{15}{00}{0.54}
  && \myCell{15}{00}{1.04} & \myCell{15}{00}{3.26}
  &&  - & - 
  && \myCell{15}{00}{0.00} & \myCell{15}{00}{0.14}\\
  \\[-0.9em]
  \multirow{3}{*}{\rotatebox{90}{$n=500$}}
  & PES
  & \myCell{40}{00}{0.00} & \myCell{40}{00}{0.01}
  && \myCell{30}{00}{0.00} & \myCell{30}{00}{0.01}
  &&  - & - 
  && \myCell{30}{00}{0.00} & \myCell{30}{00}{0.00}\\
  & PES-ES
  & \myCell{30}{00}{0.00} & \myCell{30}{00}{0.03}
  && \myCell{30}{00}{0.00} & \myCell{30}{00}{0.01}
  &&  - & - 
  && \myCell{30}{00}{0.00} & \myCell{30}{00}{0.00}\\
  & PES-2ES
  & \myCell{30}{00}{0.00} & \myCell{30}{00}{0.03}
  && \myCell{30}{00}{0.00} & \myCell{30}{00}{0.01}
  &&  - & - 
  && \myCell{30}{00}{0.00} & \myCell{30}{00}{0.00}\\
  & KPS
  & \myCell{15}{00}{0.01} & \myCell{15}{00}{0.16}
  && \myCell{15}{00}{0.62} & \myCell{15}{00}{2.39}
  &&  - & - 
  && \myCell{15}{00}{0.00} & \myCell{15}{00}{0.05}\\
  \\[-0.9em]
  \bottomrule
\end{tabular}
}
\end{table}

\newpage

\begin{table}[!ht]
\spacingset{1.1}
\centering
\tabcolsep=0.125cm 
\renewcommand{\arraystretch}{1.1}
\caption{Squared bias and variance of the estimators. Lowest/highest MSE has the darkest/lightest gray-scale. \textbf{Scenario:} No standardization of the functions in preselection step and $p=500$ grid points.}
\vspace{0.08cm}
\label{tab:sim_noscale_p500}
\resizebox{\textwidth}{!}{%
  \newcommand{\myCell}[3]{\cellcolor{black!#1}\phantom{#2}#3}
\begin{tabular}{llccccccccccccccc}
\toprule&&
  \multicolumn{2}{c}{Easy} &&
  \multicolumn{2}{c}{Complicated} &&
  \multicolumn{2}{c}{NoPoI} &&
  \multicolumn{2}{c}{OnlyPoI}&\\
  \cmidrule{3-4}
  \cmidrule{6-7}
  \cmidrule{9-10}
  \cmidrule{12-13}
  \multicolumn{2}{l}{ $\int\widehat{\beta}(t)$}&
  \multicolumn{1}{c}{Bias$^2$} & \multicolumn{1}{c}{Var.} &&
  \multicolumn{1}{c}{Bias$^2$} & \multicolumn{1}{c}{Var.} &&
  \multicolumn{1}{c}{Bias$^2$} & \multicolumn{1}{c}{Var.}&&
  \multicolumn{1}{c}{Bias$^2$} & \multicolumn{1}{c}{Var.}&\\
  \multirow{5}{*}{\rotatebox{90}{$n=250$}}
  & PES
& \myCell{15}{00}{0.09} & \myCell{15}{00}{1.18}
&& \myCell{15}{00}{1.32} & \myCell{15}{0}{11.64}
&& \myCell{15}{00}{0.00} & \myCell{15}{00}{0.02}
&& \myCell{15}{00}{0.00} & \myCell{15}{00}{0.08}\\
& PES-ES
& \myCell{20}{00}{0.06} & \myCell{20}{00}{0.94}
&& \myCell{20}{00}{1.36} & \myCell{20}{0}{10.69}
&& \myCell{40}{00}{0.00} & \myCell{40}{00}{0.01}
&& \myCell{40}{00}{0.00} & \myCell{40}{00}{0.06}\\
& PES-2ES
& \myCell{40}{00}{0.06} & \myCell{40}{000}{0.9}
&& \myCell{40}{00}{1.32} & \myCell{40}{0}{10.42}
&& \myCell{40}{00}{0.00} & \myCell{40}{00}{0.01}
&& \myCell{40}{00}{0.00} & \myCell{40}{00}{0.06}\\
& KPS
& \myCell{10}{00}{2.05} & \myCell{10}{00}{41.9}
&& \myCell{10}{0}{145.2} & \myCell{10}{}{291.28}
&& \myCell{15}{00}{0.01} & \myCell{15}{00}{0.02}
&& \myCell{10}{00}{0.04} & \myCell{10}{00}{6.06}\\
\\[-0.9em]
\multirow{5}{*}{\rotatebox{90}{$n=500$}}
& PES
& \myCell{15}{00}{0.02} & \myCell{15}{00}{0.26}
&& \myCell{15}{00}{0.48} & \myCell{15}{00}{4.67}
&& \myCell{20}{00}{0.00} & \myCell{20}{00}{0.01}
&& \myCell{20}{00}{0.00} & \myCell{20}{00}{0.01}\\
& PES-ES
& \myCell{40}{00}{0.01} & \myCell{40}{00}{0.18}
&& \myCell{20}{00}{0.46} & \myCell{20}{000}{4.1}
&& \myCell{20}{00}{0.00} & \myCell{20}{00}{0.01}
&& \myCell{20}{00}{0.00} & \myCell{20}{00}{0.01}\\
& PES-2ES
& \myCell{40}{00}{0.01} & \myCell{40}{00}{0.18}
&& \myCell{40}{00}{0.48} & \myCell{40}{00}{4.04}
&& \myCell{20}{00}{0.00} & \myCell{20}{00}{0.01}
&& \myCell{20}{00}{0.00} & \myCell{20}{00}{0.01}\\
& KPS
& \myCell{10}{00}{0.47} & \myCell{10}{0}{19.09}
&& \myCell{10}{0}{81.27} & \myCell{10}{}{229.75}
&& \myCell{10}{00}{0.01} & \myCell{10}{00}{0.01}
&& \myCell{10}{00}{0.01} & \myCell{10}{00}{2.77}\\
\\[-0.9em]
\\[-0.7em]
\multicolumn{2}{l}{ $\frac{1}{S}\sum\widehat{\beta}_s$}\\
\multirow{4}{*}{\rotatebox{90}{$n=250$}}
& PES
& \myCell{40}{00}{0.01} & \myCell{40}{00}{0.23}
&& \myCell{15}{00}{0.04} & \myCell{15}{00}{0.23}
&&  - & - 
&& \myCell{40}{00}{0.00} & \myCell{40}{00}{0.02}\\
& PES-ES
& \myCell{20}{00}{0.01} & \myCell{20}{00}{0.28}
&& \myCell{40}{00}{0.04} & \myCell{40}{00}{0.22}
&&  - & - 
&& \myCell{20}{00}{0.00} & \myCell{20}{00}{0.04}\\
& PES-2ES
& \myCell{15}{00}{0.01} & \myCell{15}{000}{0.3}
&& \myCell{40}{00}{0.04} & \myCell{40}{00}{0.22}
&&  - & - 
&& \myCell{20}{00}{0.00} & \myCell{20}{00}{0.04}\\
& KPS
& \myCell{10}{00}{0.03} & \myCell{10}{000}{0.5}
&& \myCell{10}{00}{1.17} & \myCell{10}{000}{3.2}
&&  - & - 
&& \myCell{10}{00}{0.00} & \myCell{10}{00}{0.06}\\
\\[-0.9em]
\multirow{4}{*}{\rotatebox{90}{$n=500$}}
& PES
& \myCell{15}{00}{0.00} & \myCell{15}{00}{0.13}
&& \myCell{40}{00}{0.02} & \myCell{40}{00}{0.08}
&&  - & - 
&& \myCell{20}{00}{0.00} & \myCell{20}{00}{0.00}\\
& PES-ES
& \myCell{20}{00}{0.00} & \myCell{20}{00}{0.12}
&& \myCell{40}{00}{0.02} & \myCell{40}{00}{0.07}
&&  - & - 
&& \myCell{20}{00}{0.00} & \myCell{20}{00}{0.00}\\
& PES-2ES
& \myCell{40}{00}{0.00} & \myCell{40}{00}{0.11}
&& \myCell{15}{00}{0.02} & \myCell{15}{00}{0.09}
&&  - & - 
&& \myCell{20}{00}{0.00} & \myCell{20}{00}{0.00}\\
& KPS
& \myCell{10}{00}{0.01} & \myCell{10}{000}{0.2}
&& \myCell{10}{00}{0.66} & \myCell{10}{00}{2.49}
&&  - & - 
&& \myCell{10}{00}{0.00} & \myCell{10}{00}{0.04}\\
\\[-0.9em]
\bottomrule
\end{tabular}
}
\end{table}

\newpage

\begin{table}[!ht]
\spacingset{1.1}
\centering
\tabcolsep=0.125cm 
\renewcommand{\arraystretch}{1.1} 
\caption{Mean squared bias and variance. Lowest/highest MSE has the darkest/lightest gray-scale. DGP “Complicated” with different standard deviations $\sigma_\epsilon$.}
\vspace{0.08cm}
\label{tab:sim_noise}
\resizebox{\textwidth}{!}{%
  \newcommand{\myCell}[3]{\cellcolor{black!#1}\phantom{#2}#3}
\begin{tabular}{llccccccccccccccc}
\toprule
&&\multicolumn{2}{c}{$\sigma_\epsilon=0.5$} && \multicolumn{2}{c}{$\sigma_\epsilon=1$}  && \multicolumn{2}{c}{$\sigma_\epsilon=2$} && \multicolumn{2}{c}{$\sigma_\epsilon=5$}&\\
\cmidrule{3-4}\cmidrule{6-7}\cmidrule{9-10}\cmidrule{12-13} 
\multicolumn{2}{l}{ $\int\widehat{\beta}(t)$}& \multicolumn{1}{c}{Bias$^2$} & \multicolumn{1}{c}{Var.} && \multicolumn{1}{c}{Bias$^2$} & \multicolumn{1}{c}{Var.} && \multicolumn{1}{c}{Bias$^2$} & \multicolumn{1}{c}{Var.}&& \multicolumn{1}{c}{Bias$^2$} & \multicolumn{1}{c}{Var.}&\\
\multirow{3}{*}{\rotatebox{90}{$n=250$}}
& PES
& \myCell{20}{00}{0.22} & \myCell{20}{00}{1.78}
&& \myCell{20}{00}{0.28} & \myCell{20}{00}{3.76}
&& \myCell{20}{00}{0.58} & \myCell{20}{0}{13.76}
&& \myCell{40}{00}{0.67} & \myCell{40}{0}{31.37}\\
& PES-ES
& \myCell{40}{000}{0.2} & \myCell{40}{00}{1.67}
&& \myCell{40}{00}{0.23} & \myCell{40}{00}{2.37}
&& \myCell{40}{000}{0.3} & \myCell{40}{00}{9.21}
&& \myCell{20}{00}{0.23} & \myCell{20}{0}{38.64}\\
& KPS
& \myCell{15}{0}{19.51} & \myCell{15}{0}{95.41}
&& \myCell{15}{00}{1.52} & \myCell{15}{0}{35.09}
&& \myCell{15}{00}{0.56} & \myCell{15}{0}{27.38}
&& \myCell{15}{00}{0.21} & \myCell{15}{0}{46.26}\\
\\[-0.9em]
\multirow{3}{*}{\rotatebox{90}{$n=500$}}
& PES
& \myCell{20}{00}{0.12} & \myCell{20}{00}{0.67}
&& \myCell{20}{00}{0.15} & \myCell{20}{00}{1.36}
&& \myCell{20}{00}{0.29} & \myCell{20}{00}{5.59}
&& \myCell{20}{000}{0.5} & \myCell{20}{0}{21.84}\\
& PES-ES
& \myCell{40}{00}{0.09} & \myCell{40}{00}{0.31}
&& \myCell{40}{00}{0.13} & \myCell{40}{00}{0.55}
&& \myCell{40}{000}{0.2} & \myCell{40}{000}{2.9}
&& \myCell{40}{00}{0.27} & \myCell{40}{00}{21.9}\\
& KPS
& \myCell{15}{0}{11.47} & \myCell{15}{0}{80.98}
&& \myCell{15}{000}{0.6} & \myCell{15}{0}{20.17}
&& \myCell{15}{000}{0.2} & \myCell{15}{0}{10.18}
&& \myCell{15}{00}{0.14} & \myCell{15}{0}{25.74}\\
\\[-0.9em]
\\[-0.7em]
\multicolumn{2}{l}{ $\frac{1}{S}\sum\widehat{\beta}_s$}\\
\multirow{3}{*}{\rotatebox{90}{$n=250$}}
& PES
& \myCell{40}{00}{0.01} & \myCell{40}{00}{0.05}
&& \myCell{40}{00}{0.02} & \myCell{40}{00}{0.18}
&& \myCell{20}{00}{0.03} & \myCell{20}{00}{1.04}
&& \myCell{40}{00}{2.51} & \myCell{40}{00}{9.02}\\
& PES-ES
& \myCell{40}{00}{0.01} & \myCell{40}{00}{0.05}
&& \myCell{40}{00}{0.01} & \myCell{40}{00}{0.18}
&& \myCell{40}{00}{0.01} & \myCell{40}{00}{0.69}
&& \myCell{20}{000}{1.9} & \myCell{20}{0}{11.37}\\
& KPS
& \myCell{15}{00}{0.31} & \myCell{15}{00}{2.16}
&& \myCell{15}{00}{0.17} & \myCell{15}{00}{2.15}
&& \myCell{15}{00}{0.11} & \myCell{15}{00}{3.08}
&& \myCell{15}{00}{1.79} & \myCell{15}{0}{14.06}\\
\\[-0.9em]
\multirow{3}{*}{\rotatebox{90}{$n=500$}}
& PES
& \myCell{20}{00}{0.00} & \myCell{20}{00}{0.02}
&& \myCell{20}{00}{0.01} & \myCell{20}{00}{0.06}
&& \myCell{20}{00}{0.01} & \myCell{20}{00}{0.27}
&& \myCell{20}{00}{0.97} & \myCell{20}{00}{6.06}\\
& PES-ES
& \myCell{40}{00}{0.00} & \myCell{40}{00}{0.02}
&& \myCell{40}{00}{0.01} & \myCell{40}{00}{0.05}
&& \myCell{40}{00}{0.01} & \myCell{40}{000}{0.2}
&& \myCell{40}{00}{0.42} & \myCell{40}{00}{5.86}\\
& KPS
& \myCell{15}{00}{0.19} & \myCell{15}{00}{1.61}
&& \myCell{15}{00}{0.04} & \myCell{15}{000}{0.9}
&& \myCell{15}{00}{0.03} & \myCell{15}{00000}{1}
&& \myCell{15}{00}{0.62} & \myCell{15}{00}{8.33}\\
\\[-0.9em]
\bottomrule
\end{tabular}
}
\end{table}
\vspace*{-5mm}
\begin{table}[!b]
\spacingset{1}
\centering
\tabcolsep=0.11cm
\caption{Percentage of replications with correct detection of all points of impact $\tau_1, \ldots, \tau_S$.}
\vspace{0.08cm}
\label{tab:sim_noscale_}
\tabcolsep=0.11cm
\newcommand{\myCell}[3]{\cellcolor{black!#1}\phantom{#2}#3}
 \begin{tabular}{llccccccccccccc}
\toprule
&&\multicolumn{5}{c}{300 grid points} &&& \multicolumn{5}{c}{500 grid points}\\
\cmidrule{3-7}\cmidrule{10-14}
&&\multicolumn{1}{c}{Easy} &&
\multicolumn{1}{c}{Compl.}  &&
\multicolumn{1}{c}{OnlyPoI}
&&&\multicolumn{1}{c}{Easy} &&
\multicolumn{1}{c}{Compl.}  &&
\multicolumn{1}{c}{OnlyPoI}&\\
\midrule
\multirow{4}{*}{\rotatebox{90}{$n=250$}}
& PES
& \myCell{10}{}{86.1}
&& \myCell{15}{}{28.4}
&& \myCell{15}{}{98.3}&
&& \myCell{10}{}{85}
&& \myCell{15}{}{28}
&& \myCell{10}{}{98.6}\\
& PES-ES
& \myCell{15}{}{87.4}
&& \myCell{40}{}{30.4}
&& \myCell{15}{}{98.3}&
&& \myCell{15}{}{86.5}
&& \myCell{20}{}{30.4}
&& \myCell{40}{}{98.8}\\
& PES-2ES
& \myCell{20}{}{87.5}
&& \myCell{40}{}{30.4}
&& \myCell{15}{}{98.3}&
&& \myCell{20}{}{86.6}
&& \myCell{40}{}{30.8}
&& \myCell{40}{}{98.8}\\
& KPS
& \myCell{40}{}{87.9}
&& \myCell{10}{}{25.7}
&& \myCell{40}{}{98.4}&
&& \myCell{40}{}{90.9}
&& \myCell{10}{}{24.3}
&& \myCell{15}{}{98.7}\\
\\[-0.9em]
\multirow{4}{*}{\rotatebox{90}{$n=500$}}
& PES
& \myCell{15}{}{97.6}
&& \myCell{15}{}{54.5}
&& \myCell{20}{}{100}&
&& \myCell{15}{}{97.4}
&& \myCell{15}{}{58.9}
&& \myCell{20}{}{99.9}\\
& PES-ES
& \myCell{40}{}{97.8}
&& \myCell{40}{}{56.2}
&& \myCell{20}{}{100}&
&& \myCell{40}{}{97.8}
&& \myCell{20}{}{61.1}
&& \myCell{20}{}{99.9}\\
& PES-2ES
& \myCell{40}{}{97.8}
&& \myCell{40}{}{56.2}
&& \myCell{20}{}{100}&
&& \myCell{40}{}{97.8}
&& \myCell{40}{}{61.2}
&& \myCell{20}{}{99.9}\\
& KPS
& \myCell{10}{}{95.6}
&& \myCell{10}{}{44.5}
&& \myCell{10}{}{99.8}&
&& \myCell{10}{}{96.9}
&& \myCell{10}{}{45.4}
&& \myCell{10}{}{99.4}\\
\\[-0.9em]
\bottomrule
\end{tabular}

\end{table}

\newpage

\if0\blind
{
\section*{Acknowledgements} 
The authors wish to thank Prof.~Alois Kneip (University of Bonn), Dominik Poss (University of Bonn), and Prof.~Rolf Tschernig (University of Regensburg) for their valuable suggestions that helped to improve this research work. Special thanks go to Crealytics (\url{www.crealytics.com}) for providing the data, as well as stimulating and inspiring discussions, and for posing the statistical problem considered in our application. Furthermore, we are grateful to the referees and the editors for their constructive comments which helped to improve our manuscript. 
} \fi
\if1\blind
{
\section*{Acknowledgements}
Special thanks go to Crealytics (\url{www.crealytics.com}) for providing the data, stimulating and inspiring discussions, and posing the statistical problem considered in our application.
} \fi

\section*{Supplementary Materials}
\begin{description}
\item[\textsf{R}-package and \textsf{R}-codes:] The \textsf{R}-package \texttt{FunRegPoI} contains an implementation of our estimation algorithm. The package also contains the dataset used in our real data application. The provided \textsf{R}-codes facilitate the reproduction of the results in our simulation study and our application. (supplement.zip)
\end{description}

\bibliographystyle{Chicago}
\bibliography{Manuscript.bib}

\end{document}